\documentclass[journal]{IEEEtran}

\ifCLASSINFOpdf
   \usepackage[pdftex]{graphicx}
   \graphicspath{{../pdf/}{../jpeg/}}
   \DeclareGraphicsExtensions{.pdf,.jpeg,.png}
\else
   \usepackage[dvips]{graphicx}
   \graphicspath{{../eps/}}
   \DeclareGraphicsExtensions{.eps}
\fi

\usepackage{cuted}
\usepackage{lipsum, color}

\usepackage[noadjust]{cite}
\usepackage{filecontents}

\usepackage{subfig}

\usepackage{bm}

\usepackage{amsmath}
\usepackage{amssymb}

\usepackage{epstopdf}
  
\newcommand{\rectangle}{\fboxsep0pt\fbox{\rule{1em}{0pt}\rule{0pt}{1ex}}}

\begin{document}
%
\title{Modeling the Comb Filter Effect and Interaural Coherence for Binaural Source Separation} 

\author{Luca~Remaggi,
        Philip~J.~B.~Jackson,
        Wenwu Wang,~\IEEEmembership{Senior~Member~IEEE}
\thanks{IEEE Copyright}
\thanks{The authors are with the Centre for Vision, Speech and Signal Processing, University of Surrey, Guildford, GU2 7XH, UK.  W. Wang is also with Qingdao University of Science and Technology, China. Emails: [l.remaggi,~p.jackson,~w.wang]@surrey.ac.uk.}.}

\markboth{IEEE/ACM TRANSACTIONS ON AUDIO, SPEECH, AND LANGUAGE PROCESSING, IEEE Copyright}{Remaggi \MakeLowercase{\textit{et al.}}: Modeling the Comb Filter Effect and Interaural Coherence for Binaural Source Separation} 

\maketitle

\begin{abstract}
Typical methods for binaural source separation consider only the direct sound as the target signal in a mixture. However, in most scenarios, this assumption limits the source separation performance. It is well known that the early reflections interact with the direct sound, producing acoustic effects at the listening position, e.g. the so-called comb filter effect. In this article, we propose a novel source separation model, that utilizes both the direct sound and the first early reflection information to model the comb filter effect. This is done by observing the interaural phase difference obtained from the time-frequency representation of binaural mixtures. Furthermore, a method is proposed to model the interaural coherence of the signals. Including information related to the sound multipath propagation, the performance of the proposed separation method is improved with respect to the baselines that did not use such information, as illustrated by using binaural recordings made in four rooms, having different sizes and reverberation times. 

\end{abstract}


\IEEEpeerreviewmaketitle

\section{Introduction}
Source separation is one of the most investigated fields in the signal processing community. Several application areas can benefit from it. For instance, it can improve target detection performance of passive sonar systems~\cite{SutBunSedSedFilTsiBru2010}. In biomedical engineering, source separation is often used to analyze electrocardiograms, electroencephalograms, or magnetic resonance images~\cite{UngBigStrLaz2004}. Work on ancient document restoration has utilized source separation for correcting bleed-through distortion~\cite{TonSalBed2007}.
Source separation has also been used in a large range of speech applications. For instance, it is used for improving speech enhancement~\cite{MohSmaLei2013}, crosstalk cancellation~\cite{AkeChaBulPalSumNelGat2007}, and automatic speech recognition systems~\cite{LiDenGonHae2014}. It can also be applied to improve hearing aids~\cite{HeaYoHWanWan2013}, or improve security systems~\cite{CroCriTruMur2016}. Spatial audio can also rely on it, to produce object-based audio~\cite{LiuWanJacCox2015}. Robust speech processing is another target area~\cite{KinDelGanHabHaeKelLeuMaaNakRajSehYos2016}.

In typical conditions, a sound produced by a source interacts with its environment during propagation, before it reaches a listening position. This multipath propagation is defined by its room impulse response (RIR), i.e. an acoustic signal describing the propagation of sound from source to listening position. RIRs have three parts: direct sound, early reflections, and late reverberation~\cite{Kuttruff4}. The direct sound carries information related to the source. Late reverberation provides clues about the size of the environment, without directional information~\cite{Blesser2001}. Instead, early reflections affect the human sound perception, by conveying a directional sense of the geometry of the environment~\cite{ValParSavSmiAbe2012}. This generates auditory effects, for instance modifying the source width perception~\cite{Barron1971}. Moreover, being coherent with the direct sound, strong early reflections modify the perceived sound coloration, by generating a comb filter effect~\cite{LokPatTerSilSav2011}. Hence, acoustic multipath properties should be considered in the design of source separation methods~\cite{VinBerGriBim2014}. 

Many different approaches can be found in the literature to tackle the source separation problem. However, most of them do not explicitly model the acoustic multipath properties.
For instance, in the well-known Model-based Expectation Maximization Source Separation and Localization (MESSL) method~\cite{ManWeiEll2010} only the direct sound interaural cues (i.e. the interaural phase difference (IPD) and interaural level difference (ILD)) were modeled, without considering any early reflection effect. Furthermore, although a garbage source was defined to indirectly deal with the late reverberation, there was not any formal attempt to model the reverb. 

The aim of this article is to investigate how information related to early reflections can improve source separation methods, in general. Such information can be potentially used in many source separation methods, either unsupervised or supervised. Here, we selected MESSL~\cite{ManWeiEll2010} as a baseline method due to its unsupervised nature, and the convenience in incorporating the early reflections information into its IPD model. We extended MESSL~\cite{ManWeiEll2010}, by emulating the comb filter effect produced by the early reflections. To do so, we define parametric functions in the time-frequency (TF) domain, and model the behavior of the IPD, by considering the interaction between the direct sound and the first arriving early reflection. The first reflection is chosen to be included into the model as it is the one that most affects the spatial cues~\cite{Bech1998}. Similar to MESSL, we also use an ILD model, which considers the direct sound cue, and the garbage source. 

In addition to the comb filter effect, we propose a model that separates the reverberation's effect from the rest of the RIR's. This is done by approximating the human capability of separating sounds in reverberant environments. Specifically, we model the interaural coherence (IC) of indivual sources in the mixture, similar to what was introduced in \cite{AliWanJac2013}. However, there, the target source was assumed to be in front of the listener. Here, we propose an approach that is not limited by this, but works for any target source position.       

The main novelties of this article include:
\begin{itemize}
\item a new IPD model, considering both direct sound and first reflection, to approximate the comb filter effect;
\item an extension of the MESSL IPD model, employing the target signal IC; 
\item an additional novel source separation method, obtained by combining the two new models above;
\item the application of a source and image source localization algorithm to initialize the expectation maximization (EM) algorithm used to estimate the Gaussian mixture model (GMM) parameters, and one deep-learning approach using an MLP architecture with two hidden layers to generate the TF mask.
\end{itemize}
Since the novel IPD model approximates the early reflection information, the first new pipeline is named as Early Reflection MESSL (ER-MESSL). The second novel pipeline uses the IC of the estimated target signal, hence, its name is IC-MESSL. By combining the new IPD model with the IC based model, we obtain the third proposed method, thus named as ERIC-MESSL. Finally, there is need for  the employed EM algorithm to be initialized. Since our proposed methods combine the direct sound and first reflection information, we employ our Image Source Direction and Ranging (ISDAR)~\cite{RemJacColWan2017} to initialize it, by localizing the target source and related image source~\cite{AllenBerkley79}. A comparative evaluation of early and late models is performed and reported as additional contribution. The challenging two source binaural speech mixture scenario was analyzed, by employing signal and perceptual objective measures. In the experimental section, we also evaluate the improvement given by considering early reflection information in a state-of-the-art deep learning based method, for supervised speech separation.  Through this, we further demonstrate that early reflection information improves source separation methods' performance, including deep learning, and that this can be potentially applied to many approaches in the literature.

The overall structure of this article is as follows: in Section \ref{sec:source_separation_literature}, related source separation methods are discussed; Section \ref{secchap:theoretical_definitions} defines the theoretical foundations of the proposed approach. In Sections \ref{secchap:Interaural_Cues_based_Model} and \ref{sec:precedence_effect_model}, the proposed interaural cue models for the comb filter and IC are presented, respectively. Section \ref{secchap:source_sep_model} describes the source separation algorithm. In Section \ref{secchap:exp_eval}, the experiments are described, with related results and discussion. Finally, Section \ref{secchap:conclusion6} draws the conclusion.

\section{Related Work in Speech Source Separation}
\label{sec:source_separation_literature}
Many approaches can be found in the literature to tackle the source separation problem. Some of them exploit a-priori information about basis functions representing the signals in the mixture~\cite{JanLee2003}. Others employ the non-negative matrix factorization (NMF) to learn sparse representation of speech sources~\cite{SchOls2006,ArbOzeDuoVinGriBimVan2010,JodWenEybVirSch2012,SmaFevMysMohHof2014}. The independent component analysis (ICA)~\cite{SawAraMukMak2006} is also used to decompose the mixture into independent signals, by projecting the mixtures into different domains. Scenarios where multiple microphones are available were also investigated~\cite{OzeFev2010,SouAraKinNakSaw2013,WanReiCav2016,GanVinMarGolOze2017}, e.g. using beamformers~\cite{SarKawNisLeeShi2006, DokSchVet2015}. Recently, deep neural networks (DNNs) became widely popular, when large training datasets are available~\cite{HuaKimHasJohSma2015,NugLiuVin2016,ZhaWan2016,DuTuDa2016,WanDuDai2017}.

TF masking is a popular approach, which assigns different weights to the mixture, in the TF domain~\cite{Wang2008}.
In~\cite{ManWeiEll2010}, the authors presented the MESSL method which uses binaural signals. Two interaural cues were exploited, i.e. the ILD and the IPD, relating the azimuthal sound direction of arrival (DOA) to the head orientation~\cite{HofVan1998}. The method presented in~\cite{SawAraMak2011} utilized, instead, the so called mixing vector (MV). For each frequency bin, this vector contains the time invariant frequency response component of the room. In both~\cite{ManWeiEll2010} and~\cite{SawAraMak2011}, the probability of each TF point belonging to a specific source in the mixture was determined. From this probability, TF masks were generated. In~\cite{AliJacLiuWan2014}, the two methods proposed in~\cite{ManWeiEll2010} and~\cite{SawAraMak2011} were combined, constructing a probability distribution that takes into account the three cues ILD, IPD and MV. In~\cite{DelForHor2015}, a high-dimensional vector, constructed by combining the IPD and ILD cues, was projected onto a 2D space, represented by the sound azimuth and elevation DOA. A regression approach located the sources, and estimated the TF masks. The IC cue was then employed in~\cite{HumMasBro2010}.

In the literature, yet few works can be found that consider both direct sound and early reflections. In~\cite{HuaBenChe2005}, the source separation problem was divided into different procedures, by applying deconvolution to each individual reflection. However, the performance degrades with low signal-to-noise ratio (SNR) conditions. In~\cite{NesOmo2012}, a variation of the ICA method~\cite{MakSawLee2007} was used to estimate the time-dependent mixing system, considering the multipath propagation. However, with the ICA approach, the effect of its classical permutation problem was exacerbated by the incorrect RIR components' alignment. Deconvolution of the received signals was proposed in~\cite{AsaGolBouCev2014}, by employing simulated RIRs. These RIRs were estimated by matching the temporal support of recorded ones. Nevertheless, binaural effects, such as head shadowing and pinnae influence, were not considered. Multichannel microphone arrays were used in~\cite{DokSchVet2015}, where beamformers were designed to have their directivty patterns characterized by multiple beams, to simultaneously extract direct sound and early reflections. Results show improvement with respect to classical beamforming. However, they were tested only with simulated RIRs. The work in~\cite{SchDiCDelDok2017} demonstrated the benefit of including reflection information in source separation models, by employing a NMF approach. Nevertheless, only simulated RIRs were employed.

In this article, we consider the first arriving early reflection and related direct sound, to propose a binaural model that increases the robustness in reverberant environments, by estimating TF masks. It is based on~\cite{ManWeiEll2010}, nevertheless, the proposed model could be potentially adapted to work with other methods described above, from beamformers to DNNs.

\section{Background Definitions}
\label{secchap:theoretical_definitions}
In this section, we provide a general overview of the adopted approach, and discuss the assumptions. The definitions of the general elements of the proposed architecture (e.g. binaural RIRs (BRIRs) and interaural spectrograms) are also given. 

\subsection{General Overview of the Proposed Method}
Classical source separation methods exploit features related to the direct sound to separate the target sound from a mixture. In~\cite{ManWeiEll2010}, the authors presented one of the first models to deal with the reverberation, by proposing the ``garbage'' source. In this article, we model two perceptual effects: the comb filter and IC. Through the former we aim to model the first early reflection, in a constructive fashion, to enhance the sound produced by the target speaker. The latter models the reverberation, by aiding the garbage source in suppressing it.

\begin{figure}[t]
\centering
\includegraphics[width=\columnwidth, trim={0cm 3.2cm 0cm 0cm},clip]{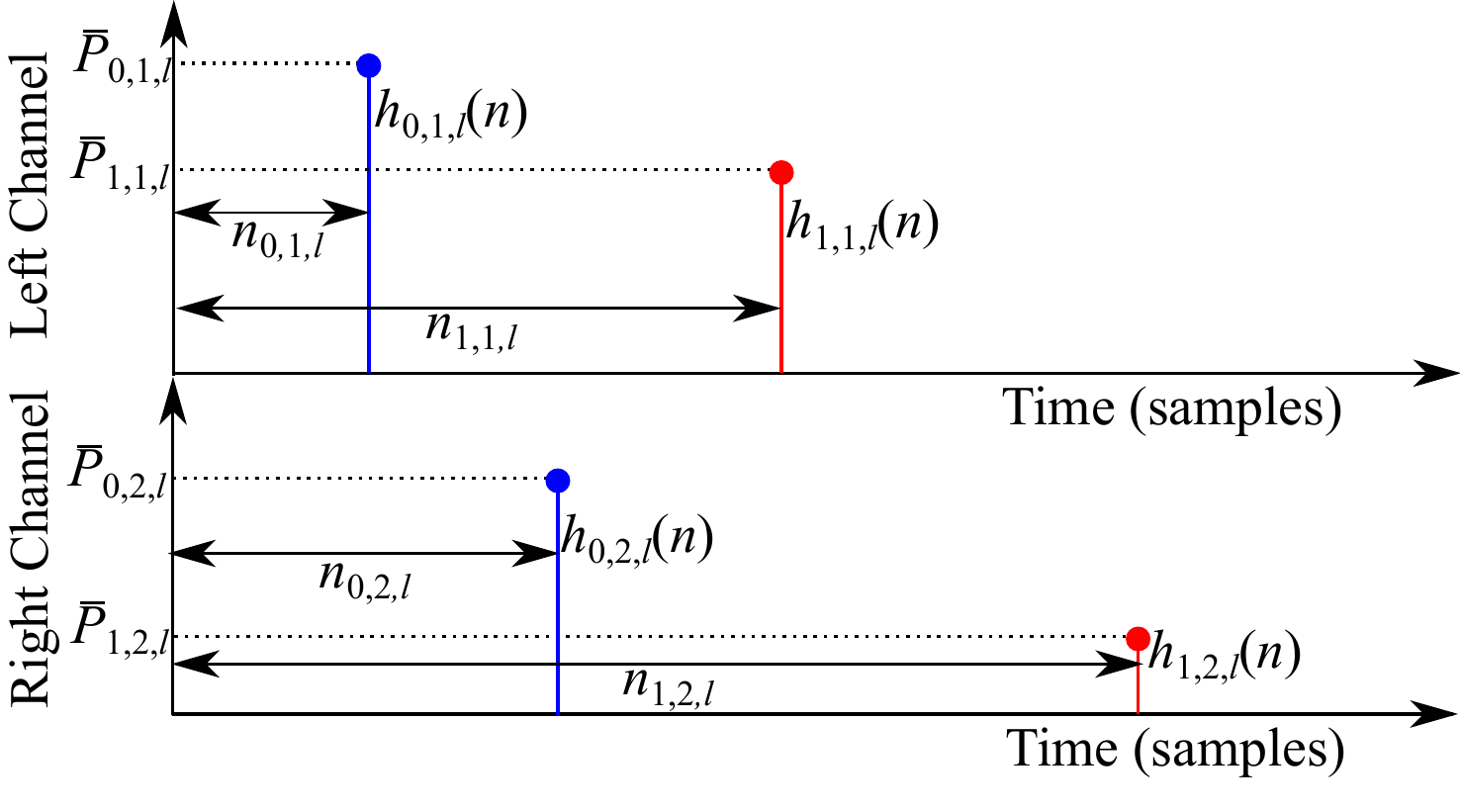}
\caption{Example of an ideal BRIR, zoomed into its direct sound (blue) and first reflection (red) components (depicted as Dirac pulses). The top figure shows the RIR related to sensor $i=1$, whereas the bottom one the RIR at sensor $i=2$. The amplitudes and delays are defined in Equation~(\ref{eq:reflections}).}
\label{fig:ideal_BRIR}
\end{figure}

\subsection{Proposed Method Assumptions}
In the proposed source separation method, assumptions were made, defining its scientific boundaries as follows:
\begin{itemize}
\item The number of sources $L$ is known a-priori;
\item Source signals are sparse in the TF domain;
\item The mixing system is time invariant;
\item The first reflection has a dominant specular component;
\item Sources are sufficiently far from the reflectors;
\item The first early reflection is coherent with the direct sound.
\end{itemize}
Although $L$ has to be known a-priori, there is no restriction on it with respect to the number of microphones $M$, thus, the method can be also applied to underdetermined scenarios. Sparsity over the TF domain corresponds to the assumption of having, for each TF bin, only one of the sources dominating the mixture. Sources and microphones are assumed to be static within a static environment, i.e. the mixing system is time invariant. Where the first reflection has a dominant specular component, it is detected from RIRs to initialize the EM re-estimation. The sources have to be distant enough from the reflectors, in order to have the first reflection arriving between 5\,ms and 40\,ms later than the direct sound. Finally, the assumption of coherence between the first reflection and direct sound allow them to be modeled as a comb filter. The later reflections, having a more stochastic nature, are assumed to be incoherent and modeled through the IC, with the reverb.

\subsection{Binaural Room Impulse Response}
\label{subsecchap:BRIR}
A RIR is a signal that characterizes the acoustics of an environment with respect to source and sensor positions. RIRs that are recorded by microphones in ear canals of a dummy head, are usually known as BRIRs. They are defined as:
\begin{equation}
I_{i,l}(n)=\sum_{e=0}^{T_m} h_{e,i,l}(n-n_{e,i,l}) + w_{i,l}(n),
\label{eq:BRIR}
\end{equation} 
where $i\in [1,2]\in \mathbb{N}$ and $l$ are the microphone and source indexes, respectively; $n$ is the discrete time index, $T_m$ indicates the last early reflection, and $w_{i,l}(n)$ represents the late reverberation, whereas $e$ is the reflection index ($e=0$ indicates the direct sound). $h_{e,i,l}$ is a function describing the reflection. $n_{e,i,l}$ represents the reflection times of arrival (TOAs). 

Following the assumption of having dominant specular components, the early reflections are approximated by Dirac deltas $\delta(n)$ of different amplitudes $\overline{P}_{e,i,l}$. 
For source separation purpose, we consider the direct sound and first reflection components (i.e. $e=\{0,1\}$) (see Fig.~\ref{fig:ideal_BRIR}):
\begin{equation}
\begin{aligned}
&h_{0,1,l}(n)=\overline{P}_{0,1,l}\delta(n-n_{0,1,l}); \\
&h_{1,1,l}(n)=\overline{P}_{1,1,l}\delta(n-n_{1,1,l}); \\
&h_{0,2,l}(n)=\overline{P}_{0,2,l}\delta(n-n_{0,2,l}); \\
&h_{1,2,l}(n)=\overline{P}_{1,2,l}\delta(n-n_{1,2,l}).
\end{aligned}
\label{eq:reflections}
\end{equation}   

\begin{figure}[t]
\centering
\includegraphics[width=0.8\columnwidth, trim={0 0 0 0},clip]{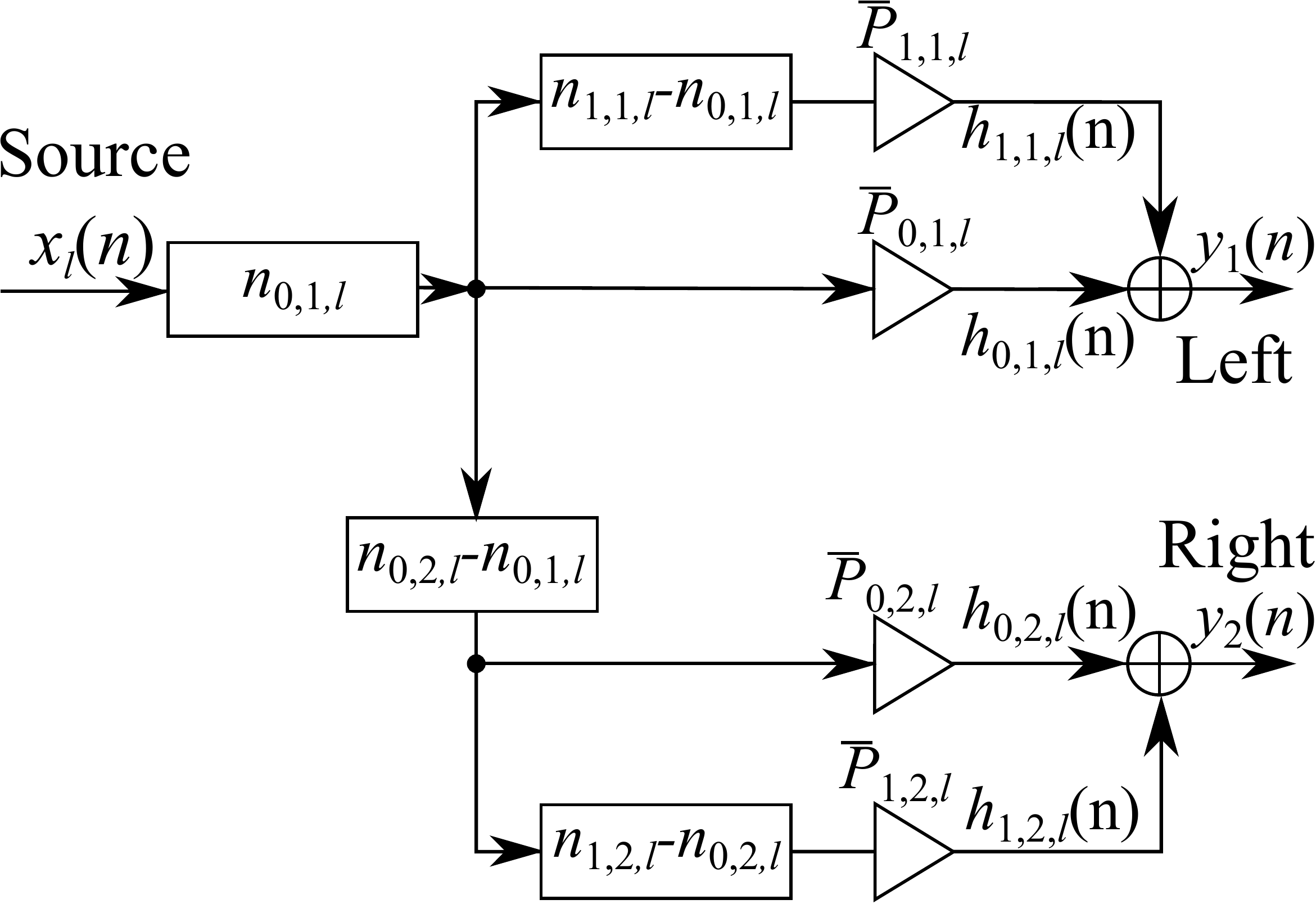}
\caption{Schematic representation of the comb filter effect created for the two received sounds ($y_1(n)$ and $y_2(n)$), given the sound produced at the $l$-th source $x_l(n)$. The direct sounds and reflections, together with the related delays (\rectangle) and attenuation factors ($\rhd$) are the same as those defined in Fig.~\ref{fig:ideal_BRIR}.}
\label{fig:comb_filter}
\end{figure}

\subsection{Comb Filter and Interaural Coherence}
\label{subsec:combandpreceffects}
In environments where the first reflection is delayed between 5\,ms and 40\,ms to the direct sound, the coloration of the sound perceived is different from the one produced~\cite{Barron1971}. In signal processing, the superimposition of a signal with its delayed version is the result of comb filtering the signal, hence, we model this perceptual effect as a comb filter effect (see Fig.~\ref{fig:comb_filter}).  

Reverberation is a diffuse component of the RIR that makes source separation more challenging by smearing the target signal, both temporally and spatially. Thus it is useful for robust separation to suppress it. With spaced microphones, reverberation signals are decorrelated above a certain frequency~\cite{VinVirGan2018}. With binaural microphones, IC measures the two signals correlation, hence we use it to model the reverberation.

\subsection{Interaural Spectrogram}
\label{subsecchap:Interaural_Spectrogram}
Following the definition of BRIR in Equation~(\ref{eq:BRIR}), the mixtures received at the $i$-th sensor can be written as:
\begin{equation}
y_i(n)=\sum_{l=1}^{L} x_l(n)* I_{i,l}(n)* w_{i,l}(n),
\label{eq:received_signal_time}
\end{equation}
where $x_l(n)$ is the signal generated by the $l$-th source, $w_{i,l}(n)$ is the convolutive white Gaussian noise, $L$ is the number of sources, and ``$*$'' is the convolution operator. 
Since the human auditory system analyzes the received mixtures in the TF domain~\cite{BroCoo1994}, we use the the short-time Fourier transform~(STFT) to calculate the TF representation of $y_i(n)$:
\begin{equation}
y_i(m,\omega)=\sum_{l=1}^L x_l(m,\omega) I_{i,l}(\omega) w_i(m,\omega),
\label{eq:received_signal_TF}
\end{equation}
where $m$ is the discrete time frame index, whereas $\omega$ is the angular frequency. $I_{i,l}(\omega)$ is not time dependent, by assuming the mixing system to be time-invariant. Considering binaural systems, the interaural spectrogram is defined as~\cite{ManWeiEll2010}:
\begin{equation}
y^{\mathrm{IS}}(m,\omega) = \frac{y_1(m,\omega)}{y_2(m,\omega)} = 10^{\alpha^{\mathrm{ILD}}(m,\omega)/20} \exp[{j\phi^{\mathrm{IPD}}(m,\omega)}],
\label{eq:interaural_spectrogram}
\end{equation}
where $\alpha^{\mathrm{ILD}}(m,\omega)$ and $\phi^{\mathrm{IPD}}(m,\omega)$ are the ILD and IPD of the observation, respectively, and $j=\sqrt{-1}$.

\section{Modeling the Comb Filter Effect}
\label{secchap:Interaural_Cues_based_Model}
The IPD and ILD cues can be modeled to generate probability distributions for identifying the dominant source, given each TF bin. The novel IPD model that approximates the comb filter effect is proposed in this section. Furthermore, the ILD model (that was presented in~\cite{ManWeiEll2010}) is described. Finally, these two are combined into a joint probability distribution. 

In the proposed model (as in MESSL~\cite{ManWeiEll2010}), sound sources are assumed to be spatially quasi-static: they have to be static within the time interval under investigation. Nonetheless, as a potential extension for future work, one could employ a tracking system, that would provide the model with updated time delays (i.e. $n_{e,i,l}$). Using audio only, beamformers could be used to estimate constantly the DOAs of the direct sound and early reflections. Alternatively, one could track sources by employing a particle filter~\cite{ValMicRou2007}, or a multimodal approach~\cite{NaqYuCha2010}.

\begin{figure}[t]
\centering
\includegraphics[width=\columnwidth, trim={0 0cm 0 0},clip]{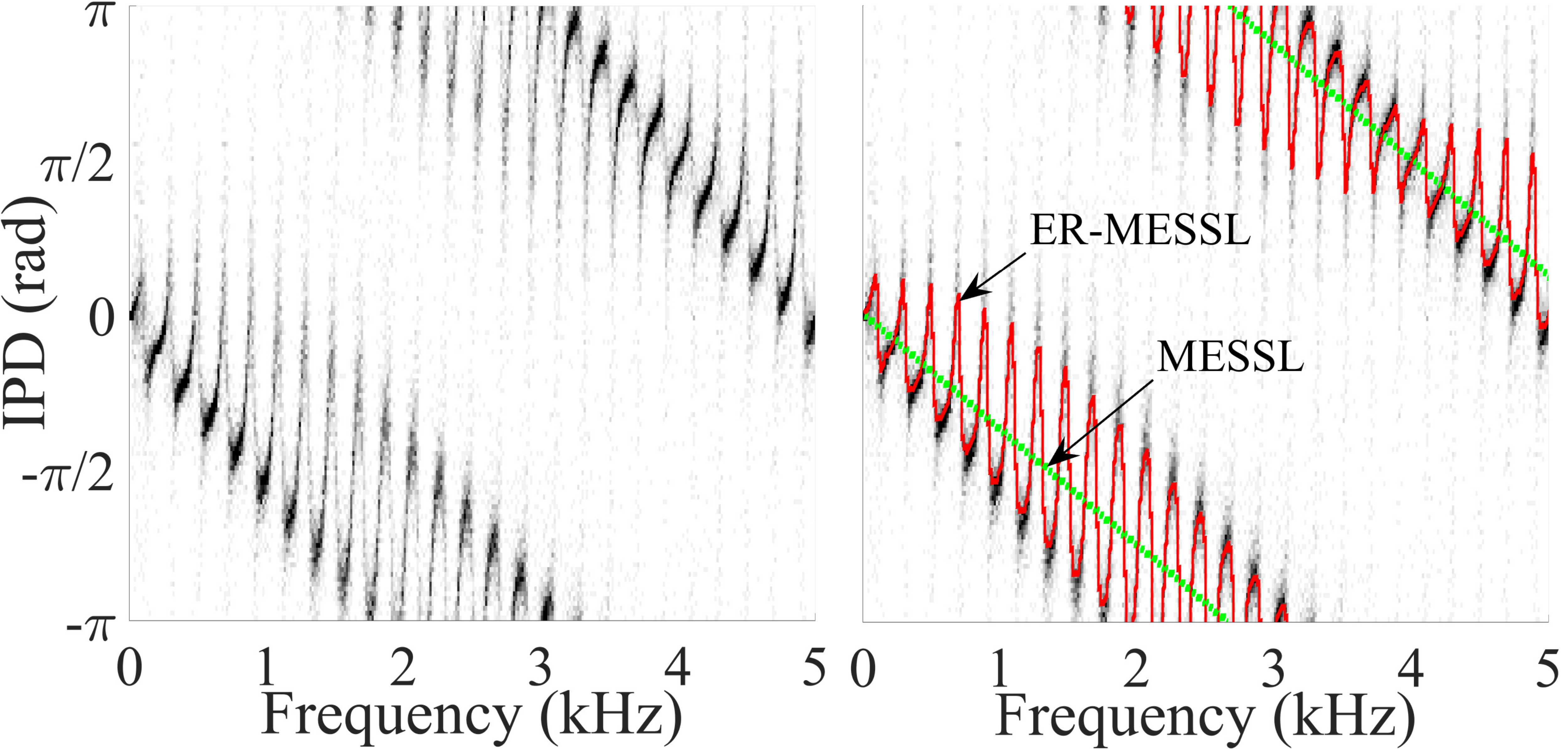}
\caption{The figure on the left shows the IPD as a function of frequency for a single source convolved with an ideal BRIR formed by only direct sound and first reflection. On the right, the same IPD function is simultaneously fitted by the MESSL IPD model \cite{ManWeiEll2010} (the straight green line), and our comb filter based ER-MESSL IPD model (the fluctuating red curve).}
\label{fig:crossphasograms}
\end{figure}

\subsection{Interaural Level and Phase Differences}
\label{subsecchap:ILD_IPD}
\begin{figure}[t]
\centering
\includegraphics[width=\columnwidth, trim={0 0cm 0 0},clip]{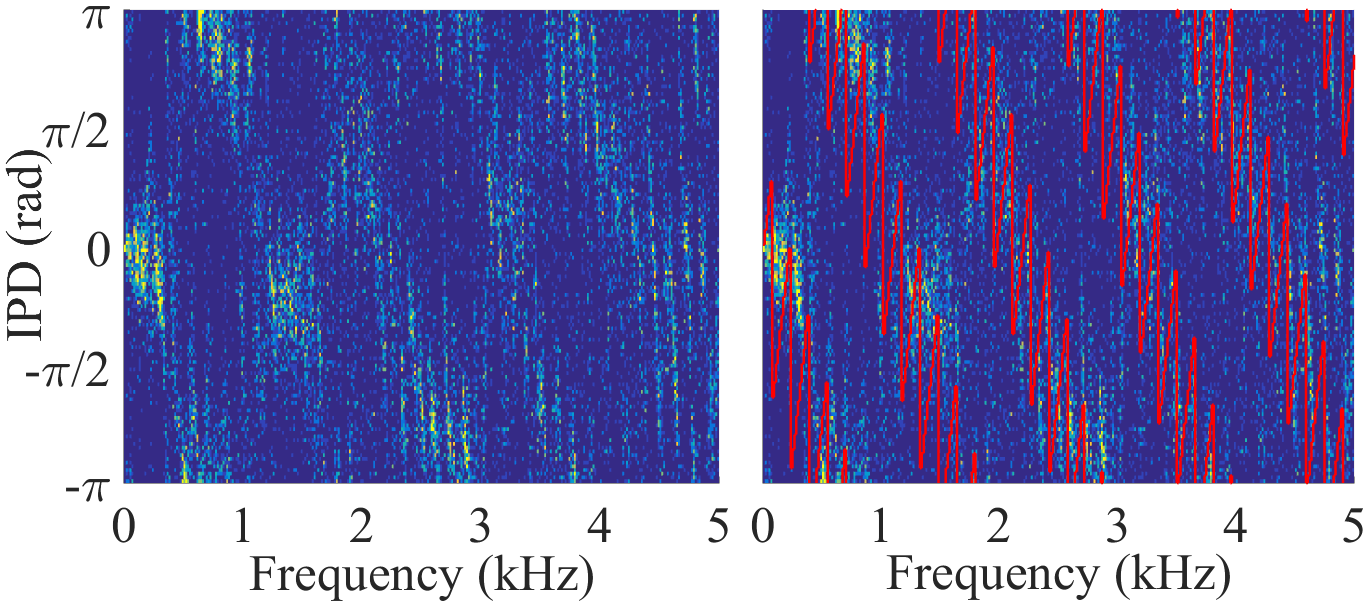}
\caption{On the left the IPD function for a mixture of two sources is shown. On the right, our comb filter based ER-MESSL IPD model (the fluctuating red curve) is employed to fit one of the two sources in the same IPD function.}
\label{fig:crossphasograms_mixtures}
\end{figure}

The proposed IPD model is defined to match the behavior of the observed IPD and is different from previous work where only the direct sound information was used~\cite{ManWeiEll2010}. By assuming ideal BRIRs as formed by direct sound and first reflection (see Fig.~\ref{fig:ideal_BRIR}), the two channel frequency responses are:
\begin{equation}
\begin{aligned}
\hat{I}_{1,l}(\omega)&= \overline{P}_{0,1,l}\exp[-j\omega n_{0,1,l}] + \overline{P}_{1,1,l} \exp[-j\omega n_{1,1,l}]); \\
\hat{I}_{2,l}(\omega)&=\overline{P}_{0,2,l}\exp[-j\omega n_{0,2,l}] + \overline{P}_{1,2,l} \exp[-j\omega n_{1,2,l}]).
\end{aligned}
\label{eq:freq_resp_channels}
\end{equation}
Their ratio is the interaural frequency response model:
\begin{equation}
\resizebox{\columnwidth}{!}{
$\begin{aligned}
&\hat{I}_{l}(\omega) =  \frac{\hat{I}_{1,l}(\omega)}{\hat{I}_{2,l}(\omega)} = \\
&\frac{\overline{P}_{0,1,l} + \overline{P}_{1,1,l} \exp[-j\omega (n_{1,1,l}-n_{0,1,l})]}{\overline{P}_{0,2,l} \exp[-j\omega (n_{0,2,l}-n_{0,1,l})] + \overline{P}_{1,2,l} \exp[{-j\omega (n_{1,2,l}-n_{0,1,l})}]}.
\end{aligned}$
}
\label{eq:interaural_freq_resp_model}
\end{equation}
The phase of this equation, denoted as $\hat{I}_{l}^{\mathrm{ang}}(\omega)$, corresponds to the proposed IPD model, and it is one of the main novelties of this article. For the $l$-th source, the difference between the observed IPD $\phi^{\mathrm{IPD}}(m,\omega)$ and its model is the phase residual:
\begin{equation}
\hat{\phi}_l^{\mathrm{IPD}}(m,\omega;\mathbf{C}_l)=\phi_l^{\mathrm{IPD}}(m,\omega)-\hat{I}_{l}^{\mathrm{ang}}(\omega;\mathbf{C}_l),
\label{eq:phase_residual}
\end{equation}
that is wrapped into the interval $[-\pi~\pi)$; and:
\begin{equation}
\mathbf{C}_l=[n_l^{\mathrm{DS}},n_l^{\mathrm{DF}},n_l^{\mathrm{ST}},\overline{P}_{0,1,l},\overline{P}_{1,1,l},\overline{P}_{0,2,l},\overline{P}_{1,2,l}],
\label{eq:model_parameters}
\end{equation}
where $n_l^{\mathrm{DS}}=n_{0,2,l}-n_{0,1,l}$, $n_l^{\mathrm{DF}}=n_{1,1,l}-n_{0,1,l}$, and $n_l^{\mathrm{ST}}=n_{1,2,l}-n_{1,1,l}$.
An example of the IPD model fitting an ideal IPD observation is shown in Fig. \ref{fig:crossphasograms}, together with a visual comparison of the MESSL IPD model \cite{ManWeiEll2010}. The ideal IPD observation was obtained from a synthetic BRIR composed of only direct sound and first reflection. From this figure, it is clear that our proposed ER-MESSL IPD model fits the observed data better than MESSL, by considering the comb filter effect. In Fig.~\ref{fig:crossphasograms_mixtures}, we also report the IPD function related to a mixture of two sources, generated using recorded BRIRs. The two sources' contributions are well visible from the figure on the left, as two linear patterns having opposite gradients. From the figure on the right, it is also visible that our proposed ER-MESSL model fits one of the two sources.

The ILD cue, $\alpha_l^{\mathrm{ILD}}(m,\omega)$, is modeled, similar to~\cite{ManWeiEll2010}, by considering directly the frequency-dependent BRIR, as:
\begin{equation}
a_l^{\mathrm{ILD}}(\omega)=20\log_{10}\left|\frac{I_{1,l}(\omega)}{I_{2,l}(\omega)}\right|,
\label{eq:ILD_model}
\end{equation}
where ``$|\cdot |$'' indicates the absolute value. 

\subsection{Interaural Cue Probability Distributions}
\label{subsecchap:PDFs}
For the ILD cue, the probability of each TF bin being associated to source $l$ can be written as a Gaussian distribution~\cite{AliJacLiuWan2014}:
\begin{equation}
p(\alpha^{\mathrm{ILD}}(m,\omega)|l)=\mathcal{N}(\alpha^{\mathrm{ILD}}(m,\omega)|\mu_l^{\mathrm{ILD}}(\omega),\sigma_l^{\mathrm{ILD}^2}(\omega)),
\label{eq:PDF_ILD}
\end{equation}
where $\mu_l^{\mathrm{ILD}}(\omega)$ is the mean, and $\sigma_l^{\mathrm{ILD}^2}(\omega)$ is the variance.

Regarding the IPD cue, a top-down approach is used to wrap the signal phase between $\pm\pi$~\cite{ManWeiEll2010}. $\hat{\phi}_l^{\mathrm{IPD}}(m,\omega;\mathbf{C}_l)$ is modeled by a Gaussian distribution:
\begin{equation}
\begin{aligned}
&p(\hat{\phi}^{\mathrm{IPD}}(m,\omega)|l,\mathbf{C}_l)= \\
&=\mathcal{N}(\hat{\phi}^{\mathrm{IPD}}(m,\omega;\mathbf{C}_l)|\mu_l^{\mathrm{IPD}}(\omega;\mathbf{C}_l),\sigma_l^{\mathrm{IPD}^2}(\omega;\mathbf{C}_l)),
\end{aligned}
\label{eq:PDF_IPD}
\end{equation}
where $\mu_l^{\mathrm{IPD}}(\omega;\mathbf{C}_l)$ and $\sigma_l^{\mathrm{IPD}^2}(\omega;\mathbf{C}_l)$ are the IPD distribution mean and variance, respectively. 

To sum up, by assuming the IPD and ILD observations as being conditionally independent given their related parameters, their probability distributions can be combined as:
\begin{equation}
\begin{aligned}
&p(\alpha^{\mathrm{ILD}}(m,\omega),\hat{\phi}^{\mathrm{IPD}}(m,\omega)|l,\mathbf{C}_l)= \\
&=\mathcal{N}(\alpha^{\mathrm{ILD}}(m,\omega),\hat{\phi}^{\mathrm{IPD}}(m,\omega;\mathbf{C}_l)|\Xi_l),
\end{aligned}
\label{eq:PDF_global}
\end{equation} 
where $\Xi_l=\{\mu_l^{\mathrm{ILD}}(\omega),\sigma_l^{\mathrm{ILD}^2}(\omega),\mu_l^{\mathrm{IPD}}(\omega;\mathbf{C}_l),\sigma_l^{\mathrm{IPD}^2}(\omega;\mathbf{C}_l)\}$.
This probability distribution identifies the proposed comb filter model, that was conceived to approximate the interaction between the received direct sound and first early reflection, i.e. two strongly coherent signals. This model does not take into account either later reflections or reverberation, which are, in this article, dealt by the IC model.   

\section{Modeling the Interaural Coherence}
\label{sec:precedence_effect_model}
To suppress reverberation, the idea is to identify those areas in the TF domain that are dominated by the direct sound, and the strong early reflections.  The direct sound and a strong reflection recorded at the two ears are highly correlated and coherent. In contrast, the late reverberation is diffuse, and does not present correlation between the binaural signals, at every frequency. Thus, we use the IC to create a probability mask, based on the coherence level, for every TF bin~\cite{AliWanJac2013}.

\subsection{Interaural Coherence TF Mask}
The process we employed to calculate the IC of a signal follows an approach that was originally proposed in~\cite{FalMer2004}, for dereverberation. For each TF bin, the auto-power spectral density of the two channels $i=\{1,2\}$ is calculated as:
\begin{equation}
\Phi_{i}(m,\omega) = \kappa \Phi_{i}(m-1,\omega) + (1-\kappa)|y_i(m,\omega)|^2,
\label{eq:APSD}
\end{equation}
where $0\leq\kappa\leq 1$ is a smoothing factor determined as $\kappa=1/(\tau\cdot f_s)$, with $\tau=10$\,ms being a time constant and $f_s$ the sampling frequency~\cite{JeuSchEscVar2010}. The cross-power spectral density between the two channels is:
\begin{equation}
\Phi_{1,2}(m,\omega) = \kappa \Phi_{1,2}(m-1,\omega) + (1-\kappa)y_1(m,\omega)y_2^*(m,\omega),
\label{eq:CPSD}
\end{equation}
with $[\cdot]^*$ indicating the complex conjugate operation. From (\ref{eq:APSD}) and (\ref{eq:CPSD}), the magnitude squared coherence is:
\begin{equation}
\Gamma_{1,2}(m,\omega) = \frac{\Phi_{1,2}(m,\omega)}{\Phi_{1}(m,\omega)\Phi_{2}(m,\omega)}.
\label{eq:MSC}
\end{equation}
The values of $\Gamma_{1,2}(m,\omega)$ are constrained between 0 and 1, thus, $\Gamma_{1,2}(m,\omega)$ is employed as the TF soft mask that models the IC. To do so, it will be used as prior mask during the posterior probability calculation, that will be described in Section~\ref{subsec:EM}\footnote{This has been implemented using the MESSL open source code's option allowing the definition of prior masks: https://github.com/mim/messl.}. $\Gamma_{1,2}(m,\omega)$ is computed from the observation by employing the equations defined in~\cite{JeuSchEscVar2010}.

The aim of modeling the IC is to suppress remaining early reflections and late reverberation, i.e. the BRIR parts that are not modeled by the comb filter.
A similar approach to calculate an IC based TF mask was employed in~\cite{AliWanJac2013}. However, there, the target source was assumed to be in front of the listener. Here, we do not make any assumption regarding the position of the target source. Its position is estimated by ISDAR, the algorithm described later, in Section~\ref{subsecchap:Initialization_Method}. Having the target source position, we then calculate $\Gamma_{1,2}(m,\omega)$ by analyzing the BRIR related to the estimated DOA. 

\subsection{The Garbage Source}
Late reflections and reverberation are problematic components of the acoustics that are undesiderable in the comb-filter model, proposed in Section~\ref{secchap:Interaural_Cues_based_Model}, as their first-order statistics are unreliable. Hence, the IC model described above is used to suppress these components of the BRIRs by consideration of their second-order statistics. In addition to this, we utilize a garbage source, as in~\cite{ManWeiEll2010}. It represents noise dominating the TF bins that are not claimed by any of the other sources. 

The parameters $\Xi^{\mathrm{G}}$ used to model the garbage source are the same as those used by the other sources to define the distribution in Equation (\ref{eq:PDF_global}). The difference is the initialization, since the garbage source is used to model the noise sources, such as background noise, measurement noise, and reverberation.

\section{Source Separation Model Reestimation}
\label{secchap:source_sep_model}
The EM is described here, along with the log-likelihood used to optimize the parameters of the proposed models.

\subsection{Parameter Estimation from Mixtures}
\label{subsecchap:Parameter_Estimation}
The parameters characterizing the interaural cue probability models are $\Omega_l=\{\Xi_l, , \beta_{l,\mathbf{C}_l}\}$, where $\beta_{l,\mathbf{C}_l}$ is the marginal class membership, described as the joint probability of each TF bin being dominated by source $l$ with the IPD model parameters $\mathbf{C}_l$: $\beta_{l,\mathbf{C}_l}=p(l,\mathbf{C}_l)$. These parameters can be estimated for a specific source $l$. This is a trivial problem upon the availability of the dominant source information for each TF bin. However, whether the source $l$ is dominating a specific TF bin is not directly observable from the mixtures. On the other hand, $l$ can be inferred from the interaural cues and observed models, that are not known a-priori. This missing data problem is solved by the EM algorithm.

The log-likelihood of the observations can be then defined as in~\cite{ManWeiEll2010}, however, with the additional IC distribution:
\begin{equation}
\resizebox{\columnwidth}{!}{
$\begin{aligned}
&\mathcal{L}(\Omega) = \sum_{m,\omega}[\log p(\alpha^{\mathrm{ILD}}(m,\omega),\hat{\phi}^{\mathrm{IPD}}(m,\omega),|\Omega) + \log\Gamma_{1,2}(m,\omega)] \\
&=\sum_{m,\omega}\log \sum_{l,\mathbf{C}_l} \beta_{l,\mathbf{C}_l} p(\alpha^{\mathrm{ILD}}(m,\omega)|l) p(\hat{\phi}^{\mathrm{IPD}}(m,\omega)|l,\mathbf{C}_l)\Gamma_{1,2}(m,\omega).
\end{aligned}$}
\label{eq:log_likelihood}
\end{equation}
This definition assumes that the IC, IPD and ILD cues are independent. As a result, the joint probability is written as the product of individual probabilities. In addition, the number of sources must be specified a-priori~\cite{ManWeiEll2010}. Note that the inclusion of the IC into the log-likelihood function is different from previous approaches, such as \cite{AliWanJac2013}. There, the IC mask was multiplied by the TF representation of the mixture. Equation~(\ref{eq:log_likelihood}) represents the proposed ERIC-MESSL. 

\subsection{Expectation-Maximization (EM)}
\label{subsec:EM}
The EM algorithm is used to estimate the parameters and probability at each TF bin. $\Gamma_{1,2}(m,\omega|l)$ is considered as a prior, and not updated during the iterations. During the E-step, the occupation likelihood of source $l$ with parameters $\mathbf{C}_l$ is calculated for each TF bin, given $\alpha^{\mathrm{ILD}}(m,\omega)$ and $\hat{\phi}^{\mathrm{IPD}}(m,\omega)$:
\begin{equation}
\begin{aligned}
\nu_l(m,\omega|\mathbf{C}_l) &= \beta_{l,\mathbf{C}_l} p(\alpha^{\mathrm{ILD}}(m,\omega)|l) \\
&\cdot p(\hat{\phi}^{\mathrm{IPD}}(m,\omega)|l,\mathbf{C}_l) p(\Gamma_{1,2}(m,\omega)|l).
\end{aligned}
\label{eq:expectation_step}
\end{equation}   

This expectation is then used in the M-step, to re-estimate the parameters, and maximize the likelihood. The ILD parameters are updated as~\cite{AliJacLiuWan2014}:
\begin{equation}
\begin{aligned}
&\mu^{\mathrm{ILD}}_l(\omega)= \frac{\sum_{m,\mathbf{C}_l}\alpha^{\mathrm{ILD}}(m,\omega)\nu_l(m,\omega|\mathbf{C}_l)}{\sum_{m,\mathbf{C}_l}\nu_l(m,\omega|\mathbf{C}_l)}, \\
&\sigma_l^{\mathrm{ILD}^2}(\omega)= \\
&\frac{\sum_{m}(\alpha^{\mathrm{ILD}}(m,\omega)-\mu_l^{\mathrm{ILD}}(\omega))^2\sum_{\mathbf{C}_l} \nu_l(m,\omega|\mathbf{C}_l)}{\sum_{m,\mathbf{C}_l}\nu_l(m,\omega|\mathbf{C}_l)},
\end{aligned}
\label{eq:ILD_maximization}
\end{equation}
whereas the IPD residual parameters are updated as:
\begin{equation}
\begin{aligned}
&\mu^{\mathrm{IPD}}_l(\omega|\mathbf{C}_l) = \frac{\sum_m\hat{\phi}_l(m,\omega|\mathbf{C}_l)\nu_l(m,\omega|\mathbf{C}_l)}{\sum_m\nu_l(m,\omega|\mathbf{C}_l)}, \\
&\sigma_l^{\mathrm{IPD}^2}(\omega|\mathbf{C}_l)= \\
&\frac{\sum_{m}(\hat{\phi}_l(m,\omega|\mathbf{C}_l)-\mu^{\mathrm{IPD}}_l(\omega|\mathbf{C}_l))^2 \nu_l(m,\omega|\mathbf{C}_l)}{\sum_{m}\nu_l(m,\omega|\mathbf{C}_l)}.
\end{aligned}
\label{eq:IPD_maximization}
\end{equation}
Also the marginal class membership is updated:
\begin{equation}
\beta_{l,\mathbf{C}_l}=\frac{1}{B}\sum_{m,\omega}\nu_l(m,\omega|\mathbf{C}_l),
\label{eq:marginal_maximization}
\end{equation}
where $B$ is the total number of TF bins.

The model parameters that are found during the last EM iteration are selected as the final estimation. Probabilistic masks are generated by marginalizing over the estimated $\mathbf{C}_l$:
\begin{equation}
M_l(m,\omega)=\sum_{\mathbf{C}_l}\nu_l(m,\omega|\mathbf{C}_l).
\label{eq:soft_masks}
\end{equation}
The separated source signal $l$ can finally be obtained as:
\begin{equation}
\hat{y}_{i,l}(m,\omega)= y_{i}(m,\omega) M_l(m,\omega),~~~~~\forall m,~\forall \omega.
\label{eq:separated_source}
\end{equation}

The seven interaural model parameters defined in $\mathbf{C}_l$ are treated in the EM as hidden variables. Specifically, they are modeled as discrete random variables, where the sets of allowed values are specified a-priori, as in~\cite{ManWeiEll2010}. The parameters in $\mathbf{C}_l$ are not internally updated by the EM algorithm. Instead, every allowed value combination is tested~\cite{ManWeiEll2010}. The combination that maximizes the log-likelihood is then chosen. 
Since the proposed IPD model in ER-MESSL and ERIC-MESSL is composed of seven parameters $\mathbf{C}_l$ (Equation~(\ref{eq:model_parameters})), it involves a seven dimensional space when trying to find the best combination of them, hence it is computationally expensive. Therefore, the amplitudes $\overline{P}_{e,i,l}$ are fixed; only the initialized value is allowed. The time-dependent parameters' allowed ranges were found empirically, as in Table \ref{tab:param_range}.

\subsection{Model Initialization}
\label{subsecchap:Initialization_Method}
\label{subsec:interaural_model_parameters}
The initialization part plays a crucial role for the EM algorithm performance, since the log-likelihood is not convex. A poor initialization leads to local maxima, thus affecting the source separation results.
The estimated source and image source positions are used to initialize the time-dependent parameters $n_l^{\mathrm{DF}}$, $n_l^{\mathrm{DS}}$ and $n_l^{\mathrm{ST}}$. Instead, the amplitudes $\overline{P}_{0,1,l}$, $\overline{P}_{1,1,l}$, $\overline{P}_{0,2,l}$, $\overline{P}_{1,2,l}$ are initialized by analyzing the BRIR that is related to the estimated DOA. Therefore, the early reflection information is not pre-estimated, but found and refined by the proposed system at each iteration. The microphone array is only used to initialize the EM algorithm.

In~\cite{ManWeiEll2010}, only the direct sound was used to model the source, and the parameters were initialized by using the GCC-PHAT algorithm~\cite{Aarabi2002}. In our proposed method, correct localization of the first reflection is also crucial. Source and image source positions are estimated through our ISDAR method~\cite{RemJacColWan2017}. This method relies on RIRs recorded via a multichannel microphone array, placed at the same listener position. We chose this since, to our knowledge, no method in the literature can reliably localize reflections, given binaural recordings. 
However, other kinds of approaches could be also employed, for instance, audio-visual based methods~\cite{KimRemJacFazHil2017}. 

ISDAR is based on spherical coordinates. Direct sound and reflection TOAs $\hat{n}_{e,i,l}$ are estimated through the clustered dynamic programming projected phase-slope algorithm (C-DYPSA), that we proposed in \cite{RemJacColWan2017}, whereas azimuth DOAs $\Theta_{e,l}$ are estimated through the delay-and-sum beamformer~\cite{VanVeenBuck1988, RemJacColWan2017}. 
Considering the listener at the center of the coordinate system, the radial distances of the source and image source are calculated as $\rho_{e,l} = \frac{1}{M}\sum_{i=1}^M (\hat{n}_{e,i,l} c_0)$, where $c_0$ is the sound speed, and $\hat{n}_{e,i,l}$ is either the estimated direct sound ($e=0$) or first reflection ($e=1$) TOA. The source and image source positions in the Cartesian coordinate system are given by $b_{x,e,l} = \rho_{e,l} \cos\Theta_{e,l}$ and $b_{y,e,l} = \rho_{e,l} \sin\Theta_{e,l}$.
Knowing the listener position, these values are converted into TDOAs to populate Equation~(\ref{eq:model_parameters}). The amplitudes $\overline{P}_{e,i,l}$ are calculated by directly analyzing the BRIRs at the reflection TOA $\hat{n}_{e,i,l}$.

Regarding the ILD distribution, the value of the ILD prior mean is estimated by utilizing a set of synthetic binaural RIRs, as in~\cite{ManWeiEll2010}. The garbage source is initialized to have a uniform distribution across IPD, and a uniform ILD distribution with zero mean for all frequencies. 

\section{Experiments and Results}
\label{secchap:exp_eval}
In this section, the results of a set of experiments are described. In these experiments, we consider mixtures of speech signals in four different recorded environments. When only the IC is modeled, and MESSL is used to model only the direct sound, the proposed method is named as IC-MESSL. When the comb filter effect is modeled, extending MESSL in that sense, without considering any prior knowledge regarding the IC, the proposed method is ER-MESSL. Otherwise, if both the comb filter and the IC are modeled, the novel method is named as ERIC-MESSL. The three proposed methods are compared to MESSL \cite{ManWeiEll2010}. The ranges of allowed parameters for the comb filter model are in Table~\ref{tab:param_range}, for each dataset. 
\begin{figure}[t]
\centering
\includegraphics[width=1\columnwidth]{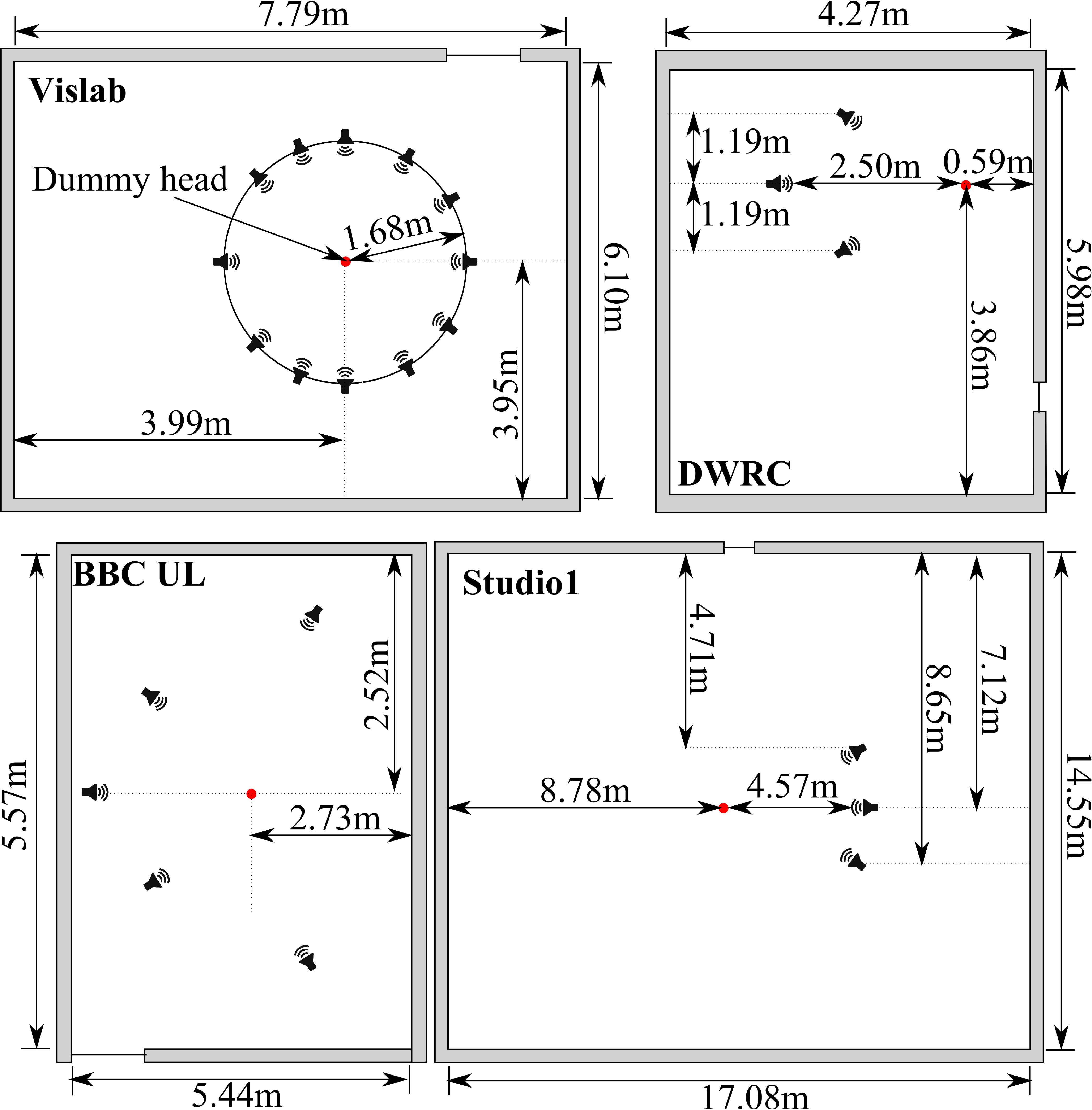}
\caption{Plan views of the four recorded rooms. The red circles represent the position of the dummy head, whereas the loudspeakers are depicted using their stylized symbol.}
\label{fig:room_pictures}
\end{figure}

At the end of this section, we also show that other separation algorithms would benefit from the inclusion of early reflection information. We extend a deep learning based state-of-the-art method. Different from MESSL, which is an unsupervised method, the deep learning approach is used to demonstrate that improvements can be achieved also for supervised methods.

\subsection{Datasets}
\label{subsecchap:datasets}
BRIRs\footnote{Available at http://cvssp.org/data/s3a, DOI: 10.15126/surreydata.00844867} were recorded in four rooms, characterized with different size and reverberation time (RT60). The four rooms are named as ``Vislab'', ``Digital World Research Centre'' (DWRC), ``BBC Usability Laboratory'' (BBC UL), and ``Studio1''. Their plan views are shown in Fig. \ref{fig:room_pictures}, whereas the RT60s are in Table \ref{tab:DRRs}, together with the number of loudspeaker positions $L_{\mathrm{TOT}}$ and their lateral angles. Two different dummy heads were employed (i.e. a Cortex Manikin Mk2 Binaural Head and Torso Simulator and a Neumann KU100 dummy head), depending on their availability for the recordings. To obtain data for the initialization, a 48-channel bi-circular array with a typical microphone spacing of 21 mm and an aperture of 212 mm was utilized to record RIRs~\cite{RemJacColWan2017}\footnote{DOIs: 10.15126/surreydata.00812228 and 10.15126/surreydata.00808465}. The dummy head and bi-circular array were recorded separately, to avoid interference effects. All the recordings were made by employing the swept-sine technique~\cite{Farina2000}, with $f_s=48$\,kHz. 

\begin{table}[!t]

\caption[]{Range sizes for the allowed values around the initialized IPD model parameters.}
\label{tab:param_range}
\centering
\resizebox{\columnwidth}{!}{

\begin{tabular}{|c|c|c|c|c|}
\cline{2-5}
\multicolumn{1}{c|}{}& \textbf{Vislab} & \textbf{DWRC} & \textbf{BBC UL} & \textbf{Studio1} \\
\hline
$\bm{n_l^{\mathrm{DF}}}$, $\bm{n_l^{\mathrm{DS}}}$, $\bm{n_l^{\mathrm{ST}}}$ & $\pm 0.13\,\mathrm{ms}$ & $\pm 0.13\,\mathrm{ms}$ & $\pm 0.19\,\mathrm{ms}$ & $\pm 0.31\,\mathrm{ms}$ \\
\hline

\end{tabular}
}
\end{table}

\begin{table}[!t]

\caption{Recorded room RT60s, averaged over the $\frac{1}{3}$ octave bands between 500\,Hz and 4\,kHz, DRRs, and TISAs, averaged over all the tested combinations. $L_{\mathrm{TOT}}$ is the number of loudspeakers. The loudspeaker positions are reported as lateral angles with respect to the dummy head orientation.}
\label{tab:DRRs}
\centering
\resizebox{\columnwidth}{!}{
\begin{tabular}{|c|c|c|c|c|}
\cline{2-5}
\multicolumn{1}{c|}{}& \textbf{Vislab} & \textbf{DWRC} & \textbf{BBC UL} & \textbf{Studio1} \\
\hline
\textbf{RT60 (s)} & $0.32$ & $0.27$ & $0.28$ & $0.94$ \\
\hline
\textbf{DRR (dB)} & $17.8$ & $3.9$ & $15.7$ & $6.0$ \\
\hline
\textbf{AVG TISA (Deg)} & $75$ & $37$ & $71$ & $32$ \\
\hline
\hline
$\bf{L^{\mathrm{TOT}}}$ & 7 & 3 & 5 & 3 \\
\hline
\textbf{Lateral angles (Deg)} & \begin{tabular}{@{}c@{}}$0, \pm30$, \\ $\pm 60, \pm 90$ \end{tabular} & $0, \pm 27$ & \begin{tabular}{@{}c@{}}$0, \pm37$, \\ $\pm 110$\end{tabular} & $0, \pm 27$ \\
\hline

\end{tabular}
}
\end{table}

\begin{figure}[t]
\centering
\includegraphics[width=\columnwidth, trim={0cm 2.9cm 0cm 0cm},clip]{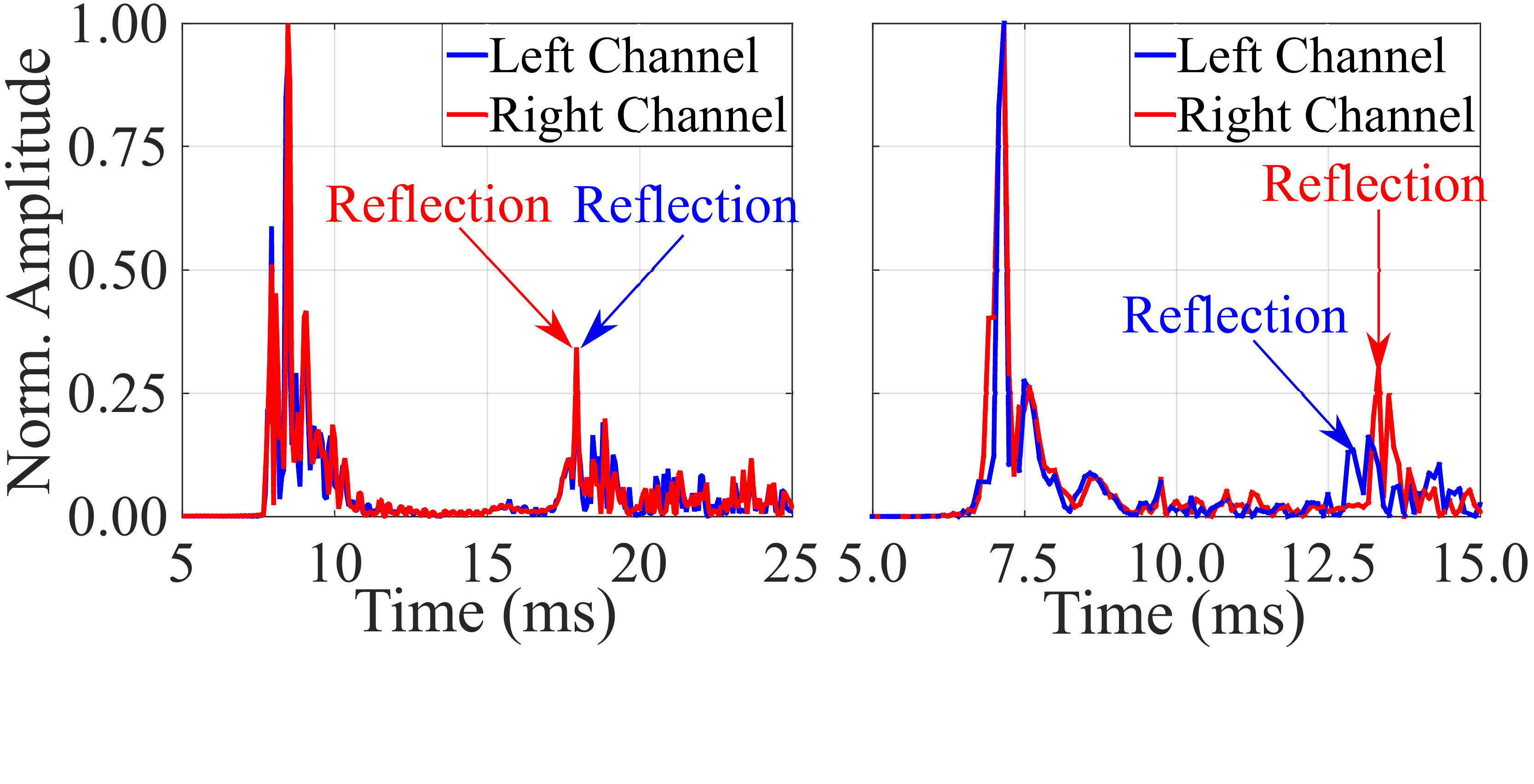}
\caption{Two BRIR absolute values, for a frontal source, zoomed into their direct sound and first reflection. On the left, reflection is generated by the floor, thus it arrives at the two ears simultaneously; on the right, reflection arrives from a lateral wall, thus there is a difference in TOAs and amplitudes.} 
\label{fig:recorded_BRIRs}
\end{figure}

\textbf{Arrangements.} Two further measures characterize the datasets: the direct to reverberant ratio (DRR) \cite{Zahorik2002}, and the average target-interferer separation angle (AVG-TISA). These will allow a more comprehensive discussion over the separation performance achieved. DRR is calculated as the ratio between the energy carried by the direct sound and the rest of the BRIR. AVG-TISA is the mean lateral angle separating the target source from the interferer, considering all the possible target-interferer combinations. DRR and AVG-TISA characterizing the four datasets are reported in Table \ref{tab:DRRs}, together with the related RT60s, and DRRs.

\textbf{Rooms.}
Vislab was an acoustically treated room at the University of Surrey, where the ``Surrey Sound Sphere'', having radius of 1.68\,m, was assembled. The loudspeakers were clamped on the sphere equator. The dummy head employed was the Cortex Manikin Mk2 Binaural Head and Torso Simulator. Both dummy head and bi-circular microphone array were placed at the sound sphere center. 

DWRC is furnished as a living room-like area. Its acoustics are representative of typical domestic living rooms. A Cortex Manikin Mk2 Binaural Head and Torso Simulator sat on a sofa. The bi-circular array was positioned right behind it. 

BBC UL is a room at the BBC R\&D center, in Salford, UK. Similar to DWRC, it is furnished to resemble a typical living room environment. A Neumann KU100 dummy head was positioned on an armchair and the bi-circular array of microphones was separately measured at the same position. 

Since the RT60s related to the three already introduced rooms were similar, an additional room was chosen: Studio1, a large recording studio at the University of Surrey. A Cortex Manikin Mk2 Binaural Head and Torso Simulator was used as dummy head. The loudspeaker positions were selected to have their height similar to the dummy head's. The microphone array was positioned about 2\,m far from the dummy head. Therefore, the image source positions found by this array were first manually modified, according to the dummy head position, before being used to initialize the EM. Depending on the loudspeaker-microphone positions in each room, reflections are generated from either the floor or lateral walls. Examples of RIRs for these two cases are depicted in Fig.~\ref{fig:recorded_BRIRs}.

\textbf{The Utterances.}
Fifteen utterances, of 3\,s length, were randomly selected from the TIMIT acoustic-phonetic continuous speech corpus~\cite{GarLamFisFisPalDah1993}. For each combination of target source and interferer(s), $U=15$ random combinations of the fifteen utterances were selected and tested. 
Therefore, the number of mixtures generated and tested for each dataset is: 
\begin{equation}
\Upsilon={L^{\mathrm{TOT}}\choose L}U,
\label{eq:number_mixtures}
\end{equation}
where the symbol ``$()$'' represents the binomial coefficient,  $L$ is the number of sources in the mixture, and $L^{\mathrm{TOT}}$ is the total number of loudspeaker positions available in the dataset. The utterances were normalized before applying the convolutions to have the same root mean square energy. 

\begin{figure*}[t] 
\vspace{-5pt}
\centering
\hspace{0cm}
\includegraphics[width=.29\textwidth, trim={11cm 3.65cm 12cm 0cm},clip]{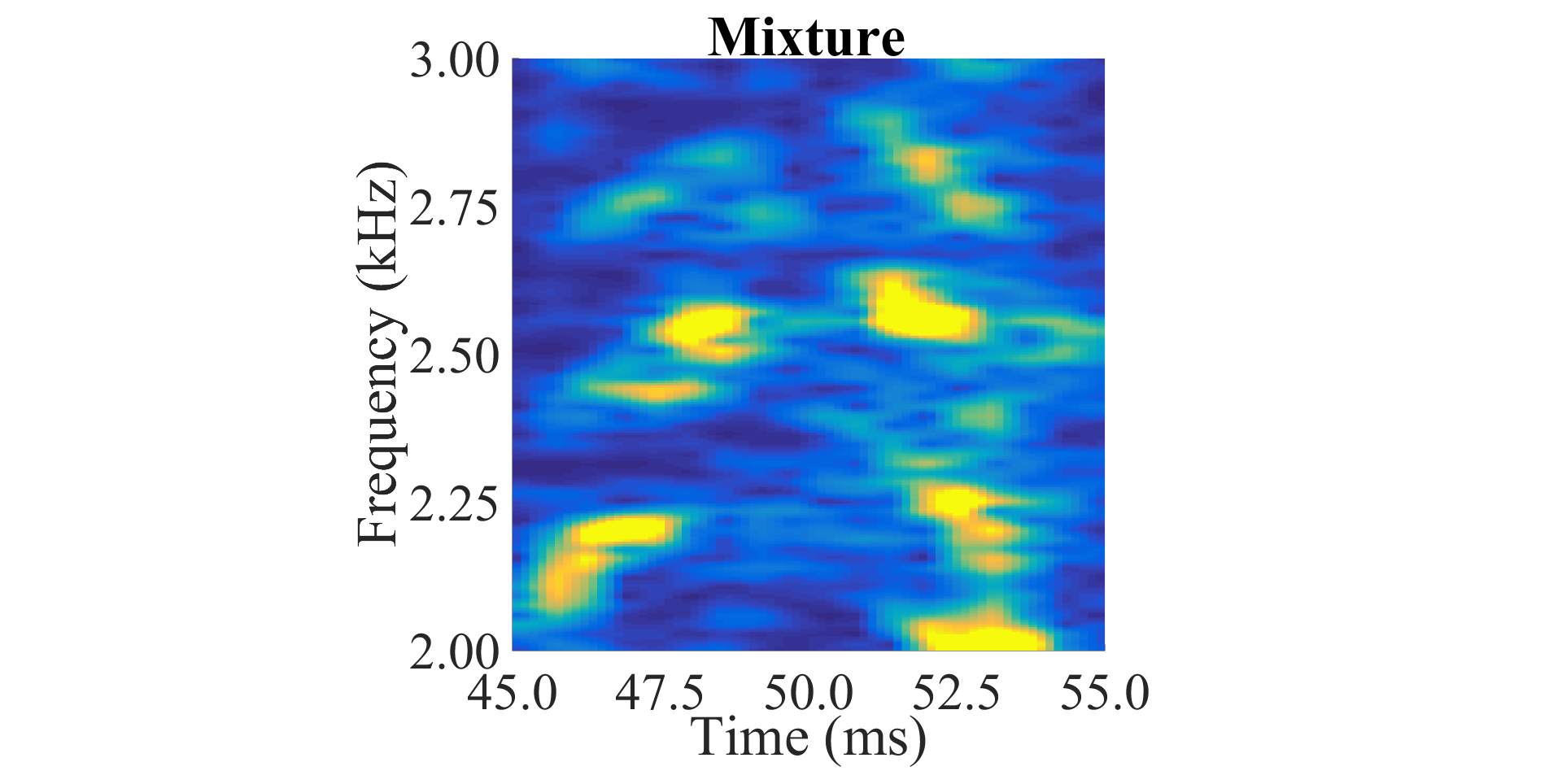}%
\label{fig:Mixture}%
\hspace{0pt}
\includegraphics[width=.29\textwidth, trim={11cm 3.65cm 12cm 0},clip]{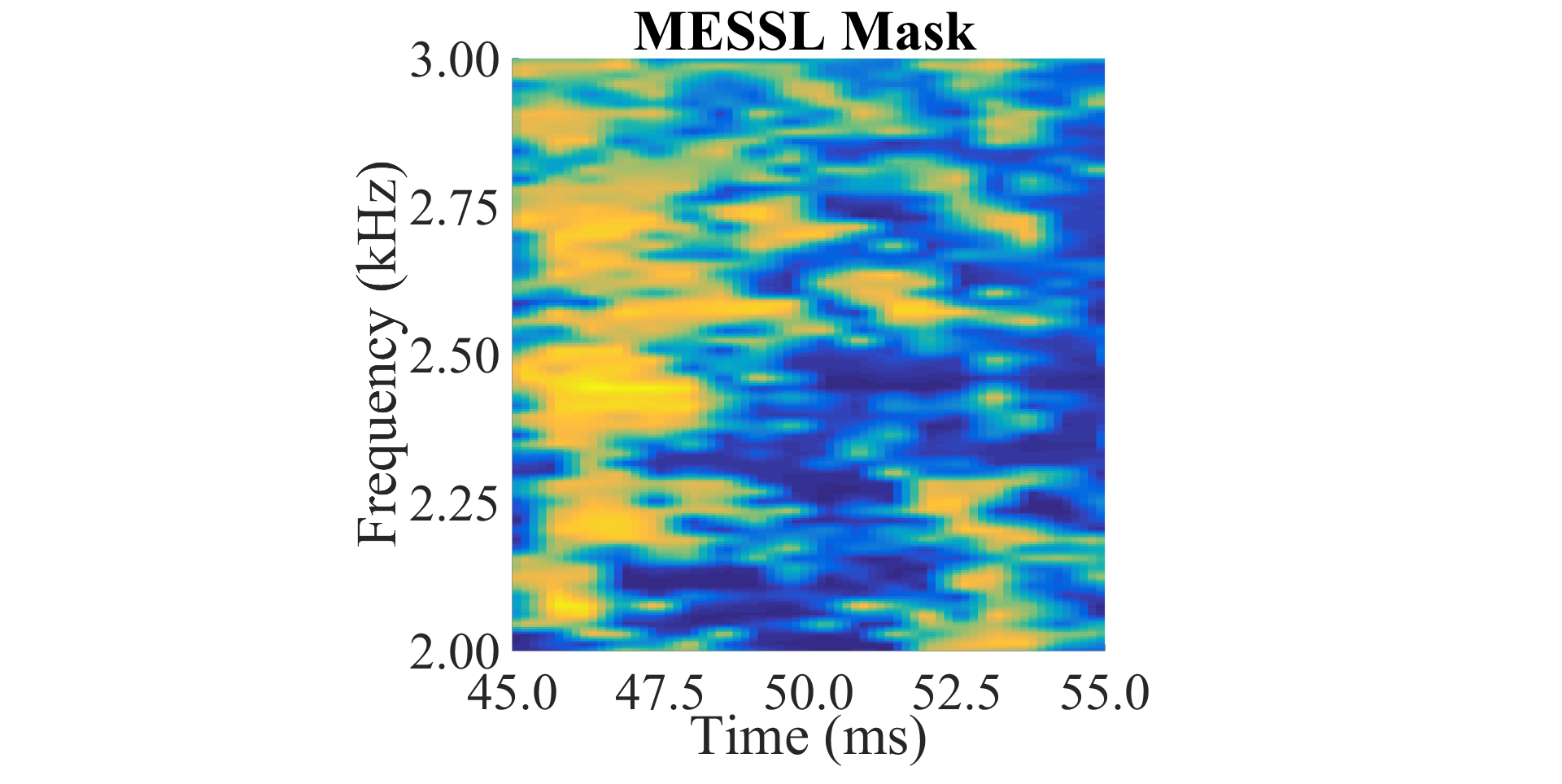}%
\label{fig:MESSL_Mask}%
\hspace{0.2cm}
\includegraphics[width=.29\textwidth, trim={11cm 3.65cm 12cm 0},clip]{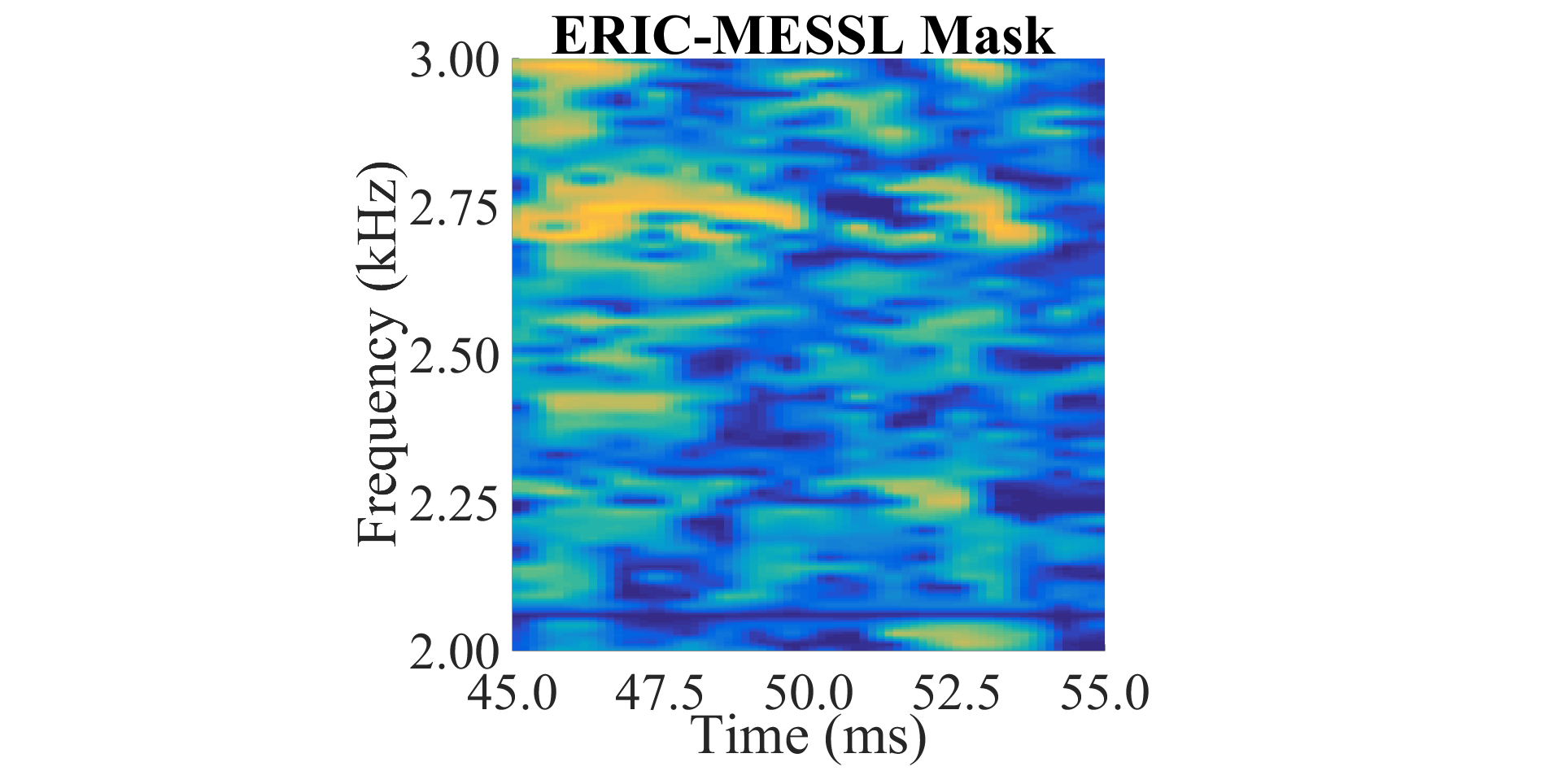}%
\label{fig:ERMESSL_Mask}%
\\
\vspace{0pt}
\hspace{0cm}
\includegraphics[width=.29\textwidth, trim={11cm 0.5cm 12cm 0},clip]{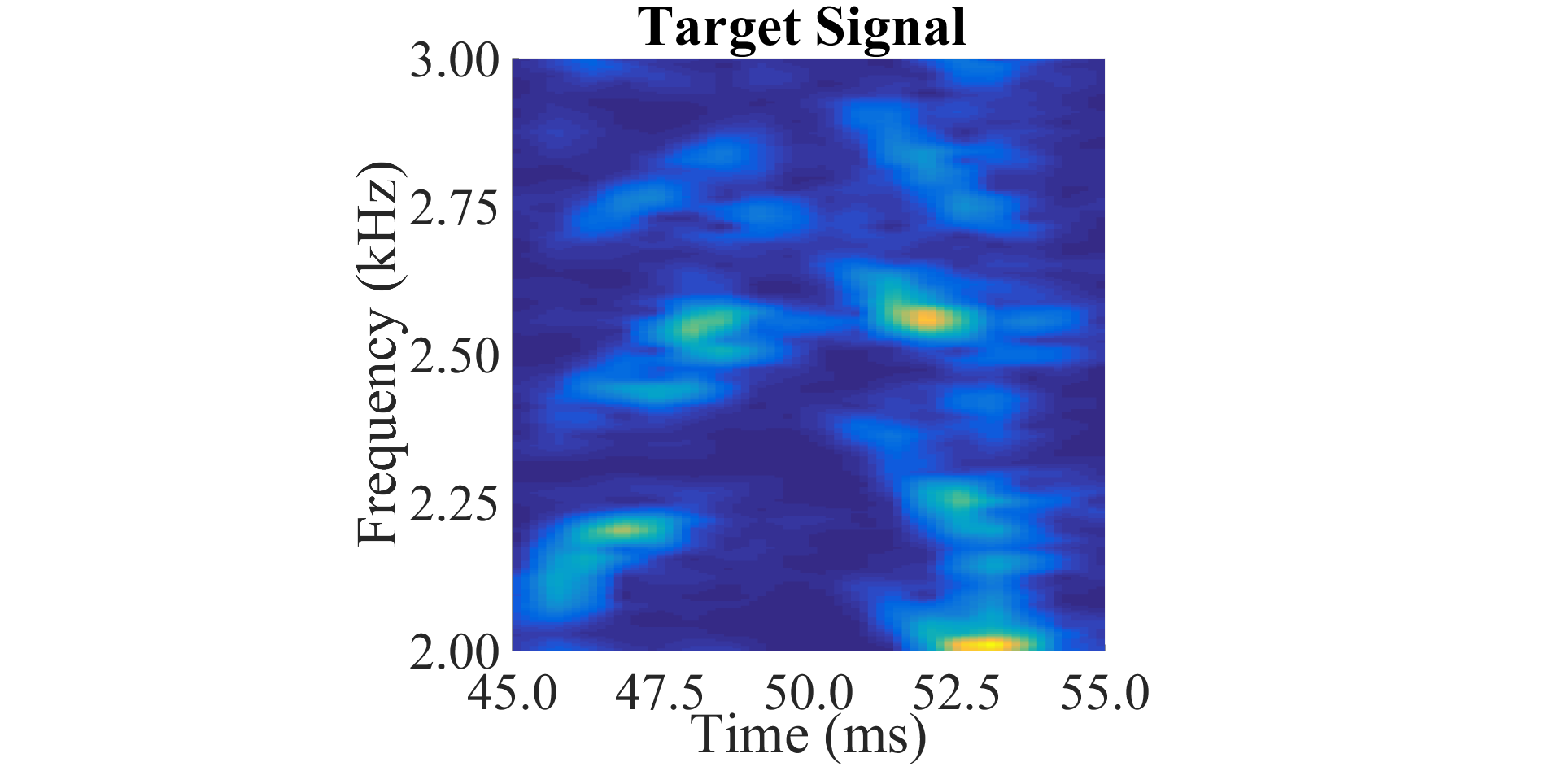}%
\label{fig:Oracle_Mask}%
\hspace{0pt}
\includegraphics[width=.29\textwidth, trim={11cm 0.5cm 12cm 0},clip]{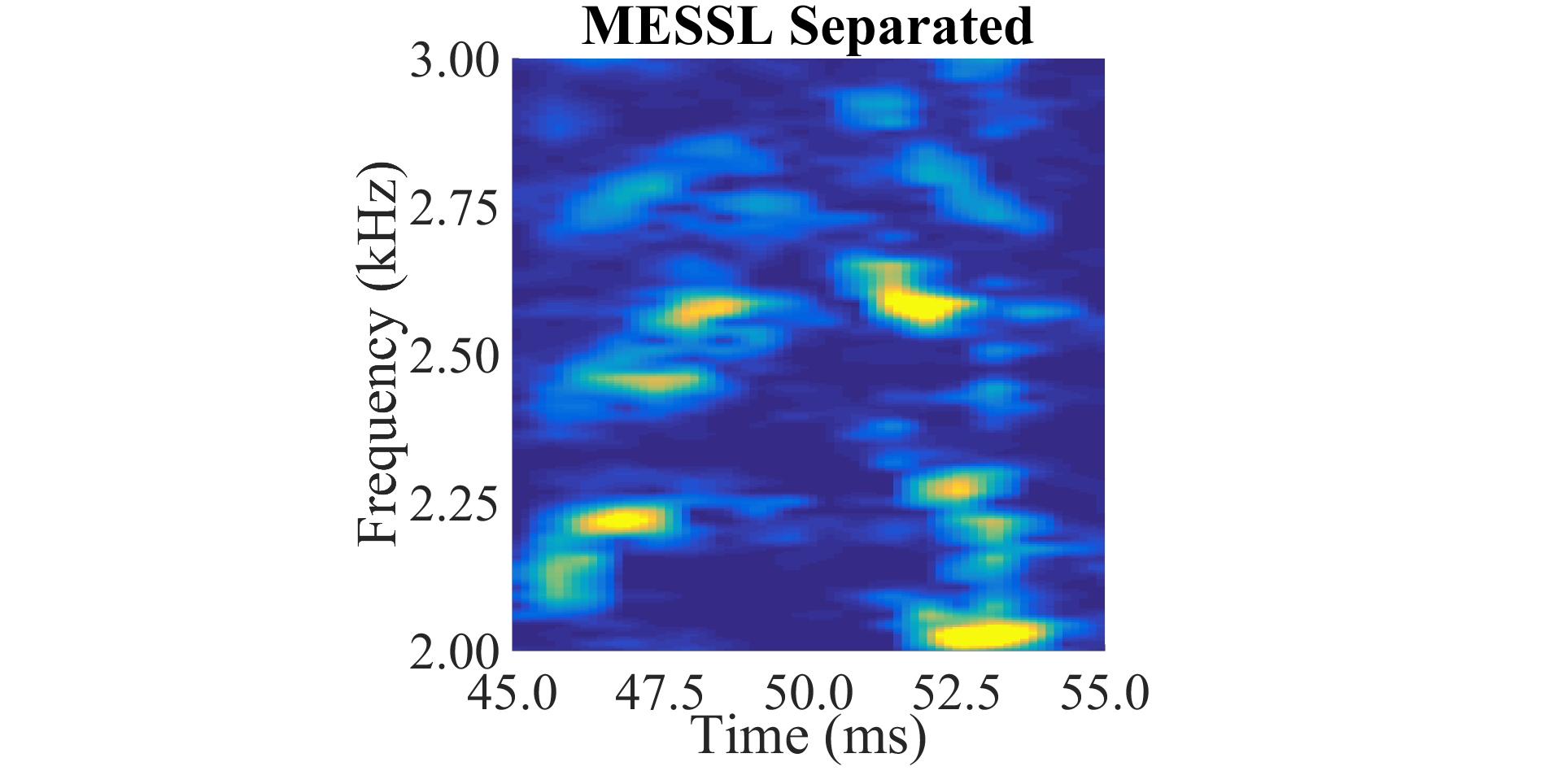}%
\label{fig:MESSLIC_Mask}%
\hspace{0.2cm}
\includegraphics[width=.29\textwidth, trim={11cm 0.5cm 12cm 0},clip]{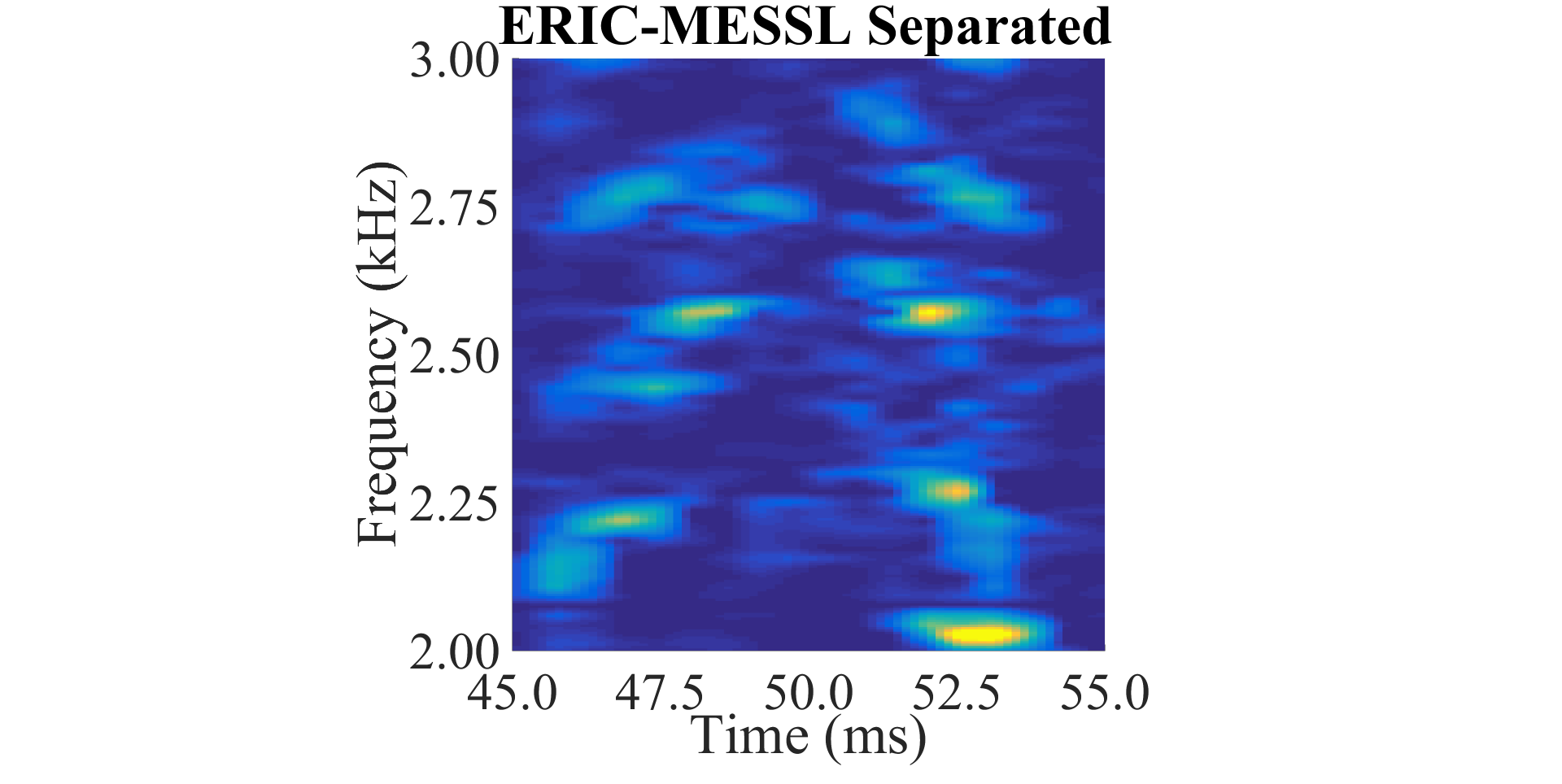}%
\label{fig:ERMESSLIC_Mask}%
\caption{The top three figures show a zoom into a mixture TF domain absolute value, the related TF masks generated by MESSL, and the TF mask estimated by the proposed ERIC-MESSL. The bottom three figures show the same TF bins of the target signal, the signal separated by MESSL, and ERIC-MESSL, respectively.}
\label{fig:TF_masks}
\end{figure*}

\subsection{Evaluation Metrics}
\label{subsecchap:eval_metrics}
The source to distortion ratio (SDR) metric is based on signal energy ratios, thus, is typically reported in dB. Following Equation~(\ref{eq:received_signal_TF}), the ideal target signal $l$, that arrives at channel $i$ free from any interference and noise, can be defined as:
\begin{equation}
y_{i,l}^{\mathrm{tar}}(m,\omega)=x_l(m,\omega)I_{i,l}(\omega).
\label{eq:target_source}
\end{equation}
Hence, the source $\hat{y}_{i,l}(m,\omega)$, separated by a source separation method as in Equation~(\ref{eq:separated_source}), can be decomposed as~\cite{VinGriFev2006}:
\begin{equation}
\hat{y}_{i,l}(m,\omega)=y_{i,l}^{\mathrm{tar}}(m,\omega)+E_{\mathrm{interf}}+E_{\mathrm{noise}}+E_{\mathrm{artif}},
\label{eq:separated_source_decomposition}
\end{equation}
where $E_{\mathrm{interf}}$ is the interference error term, $E_{\mathrm{noise}}$ the noise error term, and $E_{\mathrm{artif}}$ errors provided by general artifacts. We chose the SDR, since it emphasizes all the three error terms~\cite{VinGriFev2006}:
\begin{equation}
\mathrm{SDR}=10\log_{10}\frac{||y_{i,l}^{\mathrm{tar}}(m,\omega)||}{||E_{\mathrm{interf}}+E_{\mathrm{noise}}+E_{\mathrm{artif}}||^2},
\label{eq:SDR_calculation}
\end{equation}
where $||\cdot||$ represents the Euclidean norm operator.
Once the SDR for each of the $\Upsilon$ combinations of sources is obtained, the overall result for the dataset is calculated as their mean $\overline{\mathrm{SDR}}= \frac{1}{\Upsilon}\sum_{\upsilon=1}^{\Upsilon} \mathrm{SDR}_{\upsilon},$, where $\upsilon$ is the tested mixture index. As clean reference, we employed the target utterance convolved with the related BRIR direct sound. This is also used for the other performance metrics, described below. To extract the direct sound component from the BRIRs, we truncated them by using a Hamming window, centered at the direct sound TOA.

The perceptual evaluation of speech quality (PESQ) has been widely employed to evaluate processed speech quality~\cite{Loizou2013}. This is related to the Mean Opinion Score (MOS) of human subjective assessments, therefore, the PESQ unit of measure is MOS. Before proceeding with the PESQ value calculation, $\hat{y}_{i,l}(m,\omega)$ and $y_{i,l}^{\mathrm{tar}}(m,\omega)$ are aligned in time, in terms of amplitudes and delays, by employing Wiener filters~\cite{Loizou2013}. Through two parameters that model symmetric and asymmetric disturbances, a parametric function is then employed, mapping the differences between the processed version of $\hat{y}_{i,l}(m,\omega)$ and $y_{i,l}^{\mathrm{tar}}(m,\omega)$, to subjective assessment results~\cite{Loizou2013}. 
The overall PESQ is the mean over the $\Upsilon$ target-interferer combinations, as $\overline{\mathrm{PESQ}}= \frac{1}{\Upsilon}\sum_{\upsilon=1}^{\Upsilon}\mathrm{PESQ}_{\upsilon}$. 

Another aspect that has to be evaluated in speech signals separated via source separation algorithms is intelligibility. To do so, we employ the extended short-time objective intelligibility (ESTOI) metric~\cite{JenTaa2016}. ESTOI is a function of the separated signal $\hat{y}_{i,l}(m,\omega)$ and the clean reference $y_{i,l}^{\mathrm{tar}}(m,\omega)$. The goal of ESTOI is to produce an index (that we name as $\mathrm{ESTOI}_{\upsilon}$) that is monotonically related to the intelligibility of $\hat{y}_{i,l}(m,\omega)$~\cite{JenTaa2016}.
The overall ESTOI is the mean over the $\Upsilon$ target-interferer combinations: $\overline{\mathrm{ESTOI}}= \frac{1}{\Upsilon}\sum_{\upsilon=1}^{\Upsilon}\mathrm{ESTOI}_{\upsilon}$. 

\subsection{Control Masks}
\label{subsecchap:Oracle_mask}
Performance bounds are needed to perform a fair evaluation of source separation systems~\cite{VinAraTheNolBofSawOzeGowLutDuo2012}. Reference signals are generated from the mixtures, for comparison with the output of the proposed source separation methods. For the lower bound, random TF masks were applied to the mixture.
For the upper bound, we chose to calculate the ideal binary mask $M_l^{\mathrm{IBM}}(m,\omega)$, also known as ORACLE mask~\cite{Wang2005}. 
It is generated, for each source $l$, by comparing the $l$-th signal energy $E^{\mathrm{tar}}_l(n,\omega)$, for each TF bin, with respect to the interferers' $E^{\mathrm{int}}_{l'}(m,\omega)$ in the mixture:  
\begin{equation}
M_l^{\mathrm{IBM}}(m,\omega) = \begin{cases}
    1, & E^{\mathrm{tar}}_l(n,\omega) > E^{\mathrm{int}}_{l'}(m,\omega),~~~\forall l\ne l' \\
    0, & \text{otherwise.}
\end{cases}
\label{eq:near_optimal_mask}
\end{equation}
where $l'$ is referred to a source that is other than $l$. This equation could have also been defined by looking at the source that is louder than the sum of all other sources, instead of the loudest in general. Nevertheless, for our experiments in this article, this would not change the results, since we are focusing on cases where there are only two sources in the mixtures. 

\begin{table*}[!t]

\caption[]{$\overline{\mathrm{SDRs}}$ (left) and $\overline{\mathrm{PESQs}}$ (right) obtained by separating the target speech from a two-talker mixture.}
\label{tab:SDR_1_interferer}
\centering

\begin{tabular}{|c|c|c|c|c|c||c|c|c|c|c|c|}
\cline{2-6}\cline{8-12}
\multicolumn{1}{c|}{$\bm{\overline{\mathrm{SDR}} \mathrm{(dB)}}$}& \textbf{Vislab} & \textbf{DWRC} & \textbf{BBC UL} & \textbf{Studio1} & \textbf{AVG} & \multicolumn{1}{c|}{$\bm{\overline{\mathrm{PESQ}}  \mathrm{(MOS)}}$}& \textbf{Vislab} & \textbf{DWRC} & \textbf{BBC UL} & \textbf{Studio1} & \textbf{AVG}\\
\hline
\textbf{Random} & $-0.43$ & $-0.61$ & $-0.96$ & $0.06$ & $-0.49$ & \textbf{Random} & $1.36$ & $1.45$ & $1.45$ & $1.37$ & $1.38$ \\
\hline
\textbf{MESSL~\cite{ManWeiEll2010}} & $4.53$ & $2.54$ & $5.47$ & $0.58$ & $3.28$ & \textbf{MESSL~\cite{ManWeiEll2010}} & $1.96$ & $1.93$ & $2.06$ & $1.82$ & $1.94$ \\
\hline
\textbf{IC-MESSL} & $4.80$ & $\bm{2.73}$ & $5.79$ & $0.65$ & $3.49$ & \textbf{IC-MESSL} & $1.98$ & $\bm{1.95}$ & $\bm{2.07}$ & $\bm{1.87}$ & $1.97$ \\
\hline
\textbf{ER-MESSL} & $4.98$ & $2.68$ & $5.67$ & $0.67$ & $3.50$ & \textbf{ER-MESSL} & $2.00$ & $1.93$ & $2.06$ & $1.83$ & $1.96$ \\
\hline
\textbf{ERIC-MESSL} & $\bm{5.14}$ & $2.70$ & $\bm{5.89}$ & $\bm{0.75}$ & $\bm{3.62}$ & \textbf{ERIC-MESSL} & $\bm{2.01}$ & $\bm{1.95}$ & $\bm{2.07}$ & $\bm{1.87}$ & $\bm{1.98}$ \\
\hline
\textbf{ORACLE} & $6.21$ & $5.04$ & $6.82$ & $0.88$ & $4.66$ & \textbf{ORACLE} & $2.34$ & $2.45$ & $2.45$ & $1.96$ & $2.30$ \\
\hline

\end{tabular}
\end{table*}

\begin{table}[!t]

\caption[]{$\overline{\mathrm{ESTOIs}}$ obtained by separating the target speech from a two-talker mixture.}
\label{tab:ESTOI_1_interferer}
\centering

\begin{tabular}{|c|c|c|c|c|c|}
\cline{2-6}
\multicolumn{1}{c|}{$\bm{\overline{\mathrm{ESTOI}}}$}& \textbf{Vislab} & \textbf{DWRC} & \textbf{BBC UL} & \textbf{Studio1} & \textbf{AVG} \\
\hline
\textbf{Random} & $0.19$ & $0.17$ & $0.19$ & $0.05$ & $0.15$ \\
\hline
\textbf{MESSL~\cite{ManWeiEll2010}} & $0.28$ & $0.22$ & $0.30$ & $0.07$ & $0.22$ \\
\hline
\textbf{IC-MESSL} & $0.29$ & $0.23$ & $0.31$ & $0.07$ & $0.23$ \\
\hline
\textbf{ER-MESSL} & $0.29$ & $0.23$ & $0.30$ & $0.08$ & $0.23$ \\
\hline
\textbf{ERIC-MESSL} & $\bm{0.29}$ & $\bm{0.24}$ & $\bm{0.31}$ & $\bm{0.10}$ & $\bm{0.24}$ \\
\hline
\textbf{ORACLE} & $0.34$ & $0.29$ & $0.36$ & $0.10$ & $0.27$ \\
\hline

\end{tabular}
\end{table}

\begin{table}[!t]

\caption[]{P-values obtained from a paired t-test that compared the SDRs using MESSL, with the SDRs using each of the three proposed methods.}
\label{tab:ttest}
\centering

\begin{tabular}{|c|c|c|c|c|c|}
\cline{2-6}
\multicolumn{1}{c|}{}& \textbf{Vislab} & \textbf{DWRC} & \textbf{BBC UL} & \textbf{Studio1} & \textbf{AVG} \\
\hline
\textbf{IC-MESSL} & $\bm{0.0}\,\textbf{\%}$ & $\bm{0.0}\,\textbf{\%}$ & $\bm{0.0}\,\textbf{\%}$ & $7.9\,\%$ & $\bm{0.0}\,\textbf{\%}$ \\
\hline
\textbf{ER-MESSL} & $\bm{0.0}\,\textbf{\%}$ & $8.6\,\%$ & $\bm{0.0}\,\textbf{\%}$ & $12.0\,\%$ & $\bm{0.0}\,\textbf{\%}$ \\
\hline
\textbf{ERIC-MESSL} & $\bm{0.0}\,\textbf{\%}$ & $68.9\,\%$ & $\bm{0.0}\,\textbf{\%}$ & $\bm{4.1}\,\textbf{\%}$ & $\bm{0.0}\,\textbf{\%}$ \\
\hline

\end{tabular}
\end{table}

\subsection{Source Separation Experiments}
The experiments performed were focused on analyzing the source separation performance, employing mixtures composed of two sources ($L=2$), i.e. target and interferer. These experiments were designed to compare our three novel methods (i.e. IC-MESSL, ER-MESSL and ERIC-MESSL) with the baseline (i.e. MESSL~\cite{ManWeiEll2010}), that models only the direct sound IPD, by calculating the $\overline{\mathrm{SDR}}$ and $\overline{\mathrm{PESQ}}$ scores. Results obtained by applying the ideal masks are also reported as reference. 

\begin{figure}[!t]
\centering
\includegraphics[width=1\columnwidth, trim={0cm 3.2cm 0.2cm 0.52cm},clip]{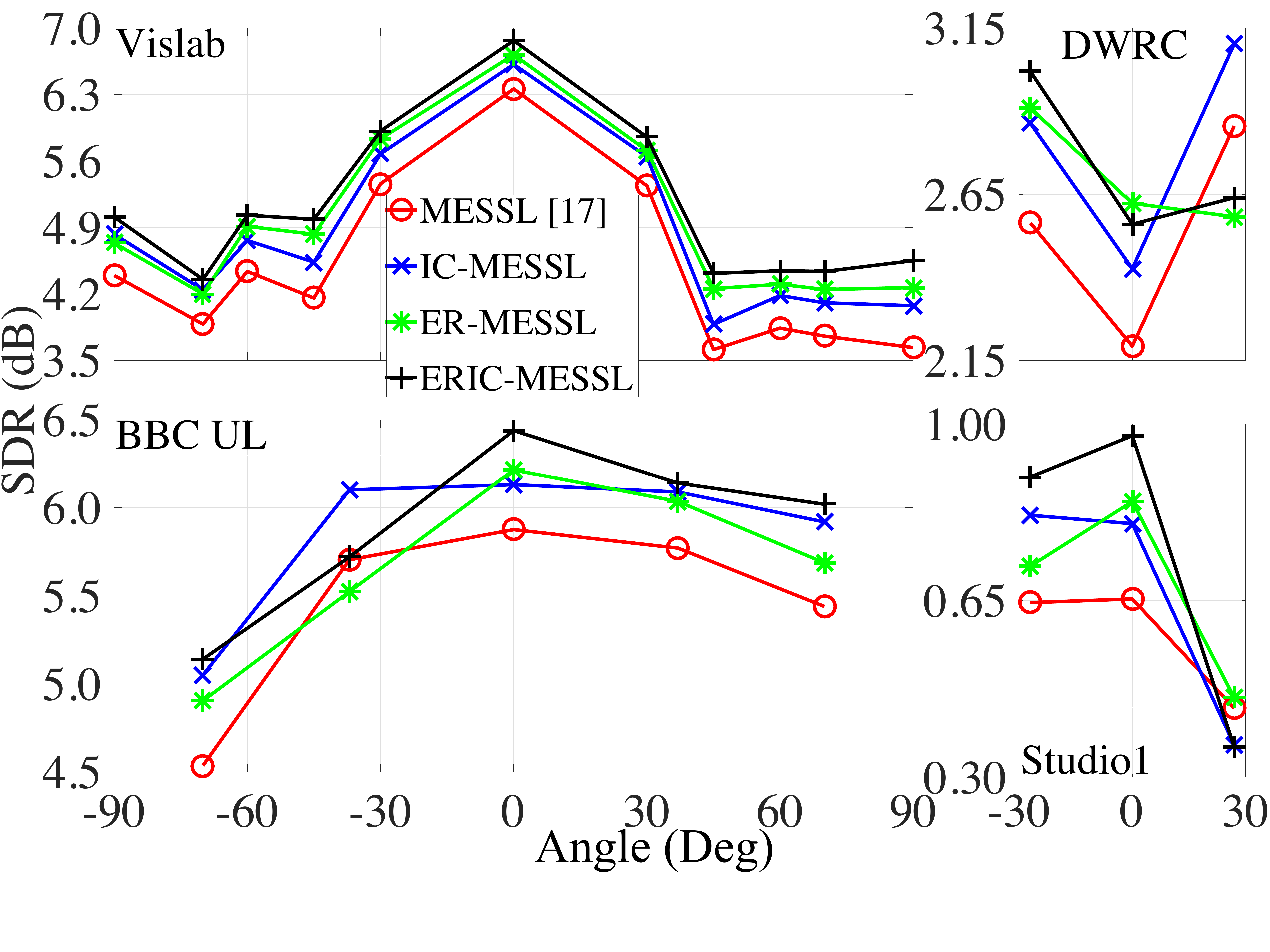}
\caption{SDRs obtained by separating a target speech from a two-talker mixture. These results refer to different target source positions, averaged over every interferer position. }
\label{fig:angle_datasets}
\end{figure}

The number of maximum iterations for the EM algorithm was set, for all the experiments, to be 16. The smoothing factor to calculate the IC was set to be $\kappa=0.5$. The BRIRs and the utterances introduced in Section~\ref{subsecchap:datasets} were utilized to create the reverberant mixtures described in Equation~(\ref{eq:received_signal_time}). 
Since the BRIRs were recorded having, within the same dataset, constant distance between loudspeakers and listening position, the target-to-interferer ratio (TIR) in the mixture was equal to 0\,dB. This choice was made to focus the evaluation on the source separation methods' performance, by avoiding their dependency on the variation in utterance energy and source distance. Furthermore, TIR equal to 0\,dB represents a challenging case, where no distinction can be made between target and interferer by looking at their energy levels.  

Examples of masks generated by MESSL and the proposed ER-MESSL are depicted in Fig.~\ref{fig:TF_masks}. We can observe that differences between the two masks are pronounced. These differences lead to the TF representation of the signal separated through ERIC-MESSL to be more similar to the groundtruth target signal, when compared to MESSL's separated signal. 

For our experiments we used the open-source code of MESSL, where we set to the frequency-dependent parameter modeling option. The tested MESSL model, hence, includes a non-parametric modeling of the ``impurities'' around the direct sound component. Nevertheless, in MESSL, the early reflection model was not directly defined through parameters. Instead, we drive our system to extract the information related to both direct sound and early reflection. We also use the frequency-dependent parameter modelling (pre-implemented in MESSL) to model the impurities around the estimation.

\subsection{Source Separation Results}
The SDR side of Table \ref{tab:SDR_1_interferer} shows that ERIC-MESSL, the proposed source separation method that models both the comb filter and IC, outperforms the baseline (i.e. the MESSL method \cite{ManWeiEll2010}), when applied to any of the four datasets. Furthermore, it provides better performance if compared to the other proposed methods. However, for the DWRC dataset, the other proposed method IC-MESSL produces the highest SDR. This is due to strong reflections arriving from different directions with respect to the direct sound, which corresponds to a lower impact of the comb filter effect \cite{LokPatTerSilSav2011}. Observing PESQ in Table \ref{tab:SDR_1_interferer}, in general, the two proposed methods that model the IC (i.e. IC-MESSL and ERIC-MESSL) have comparable results, and are both better than the other methods. However, in acoustically controlled environments, such as Vislab, the first reflection direction is initialized more accurately by ISDAR, and the comb filter model performs better, with ERIC-MESSL having a higher PESQ. This shows the importance of an accurate initialization of the GMM parameters. Similar trends are reported in Table~\ref{tab:ESTOI_1_interferer}, where the ESTOIs related to the proposed methods are greater than the baseline. ESTOI results show ERIC-MESSL to be the best proposed method, providing a greater intelligibility for every dataset.

In general, DWRC and Studio1 are more challenging datasets, producing low SDR, PESQ and ESTOI values for every tested method. The reason can be found in Table \ref{tab:DRRs}: they have low DRRs and narrow AVG-TISAs. Low DRR entails difficulties for each of the algorithms, since the IPD curve, that was described in Fig. \ref{fig:crossphasograms}, is highly distorted by the strong reverberation. At the same time, narrow AVG-TISA affects the overall results, since small angles between target and interferer correspond to small variations between the IPD and ILD cues related to the two signals in the mixture.   

\begin{table*}[!t]

\caption[]{$\overline{\mathrm{SDRs}}$ (left) and $\overline{\mathrm{PESQs}}$ (right) obtained by separating the target speech from a two-talker mixture. These results are calculated by considering only recording setups where direct sound and first reflection have same DOA.}
\label{tab:SDR_1_interferer_same_angle}
\centering

\begin{tabular}{|c|c|c|c|c||c|c|c|c|c|}
\cline{2-5}\cline{7-10}
\multicolumn{1}{c|}{$\bm{\overline{\mathrm{SDR}} \mathrm{(dB)}}$}& \textbf{DWRC} & \textbf{BBC UL} & \textbf{Studio1} & \textbf{AVG} & \multicolumn{1}{c|}{$\bm{\overline{\mathrm{PESQ}}  \mathrm{(MOS)}}$} & \textbf{DWRC} & \textbf{BBC UL} & \textbf{Studio1} & \textbf{AVG} \\
\hline
\textbf{MESSL~\cite{ManWeiEll2010}} & $2.00$ & $5.22$ & $0.55$ & $2.59$ & \textbf{MESSL~\cite{ManWeiEll2010}} & $1.86$ & $2.04$ & $1.87$ & $1.92$ \\
\hline
\textbf{IC-MESSL} & $2.26$ & $5.57$ & $0.68$ & $2.84$ & \textbf{IC-MESSL} & $\bm{1.88}$ & $2.05$ & $2.92$ & $1.95$ \\
\hline
\textbf{ER-MESSL} & $2.43$ & $5.60$ & $0.80$ & $2.94$ & \textbf{ER-MESSL} & $1.86$ & $2.06$ & $1.92$ & $1.95$ \\
\hline
\textbf{ERIC-MESSL} & $\bm{2.70}$ & $\bm{5.80}$ & $\bm{0.87}$ & $\bm{3.12}$ & \textbf{ERIC-MESSL} & $\bm{1.88}$ & $\bm{2.07}$ & $\bm{1.95}$ & $\bm{1.97}$ \\
\hline

\end{tabular}
\end{table*}

Assuming the $\Upsilon$ SDR results of each dataset as being normally distributed, the paired t-test was performed to determine whether the results, generated through the three proposed methods, are significantly different from the ones obtained by MESSL. In Table~\ref{tab:ttest}, the p-values are reported. They represent the probability of rejecting the hypothesis that the two sets under investigation are statistically different (i.e. a low p-value means that the two sets are statistically different). By looking at the results averaged over all the datasets by comparing every tested sample, with a significance level of 5\,\%, we can state that the results of IC-MESSL, ER-MESSL, and ERIC-MESSL are statistically different from those of MESSL. Moreover, by looking at each dataset singularly, results show that the three proposed methods are statistically different from MESSL in Vislab and BBC UL. However, in DWRC and Studio1 this is valid only for IC-MESSL and ERIC-MESSL, respectively. These results confirm what was already shown in Table~\ref{tab:SDR_1_interferer}, where the improvement given by IC-MESSL, ER-MESSL, and ERIC-MESSL is, in general, higher in BBC UL and Vislab than in DWRC and Studio1. The statistical significance of the results demonstrates the key point of the manuscript, which is about the importance of considering early reflection information when constructing a source separation model. 

\begin{figure}[!t]
\centering
\includegraphics[width=\columnwidth, trim={0cm 1.45cm 0.2cm 0.65cm},clip]{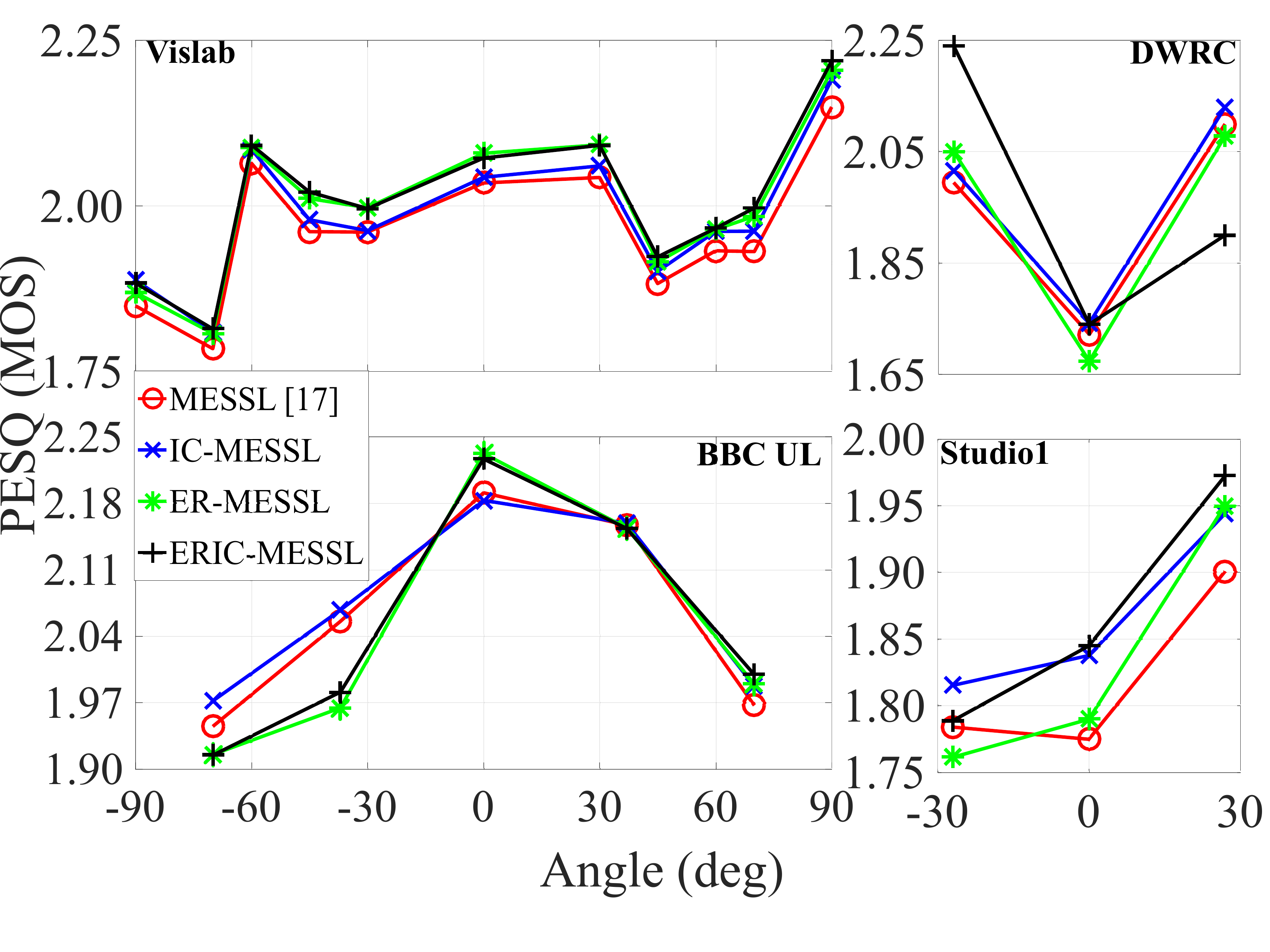} 
\caption{PESQs obtained by separating a target speech from a two-talker mixture. These results refer to different target source positions, averaged over every interferer position.}
\label{fig:angle_datasets_PESQ}
\end{figure}

For the four datasets, the SDR results can also be reported as a function of the target source location, as shown in Fig.~\ref{fig:angle_datasets}. For each target source position, within the dataset, the SDR is calculated by considering each of the correspondent interferer locations. Then, the obtained SDRs are averaged over these interferer positions, leading to one result for each target source location. Due to the  cone of confusion, which is well-known for IPD based localization methods \cite{WenArrKisWig1993}, it is not possible to discriminate between the IPD of two sources lying at the same lateral angle. Therefore, results are reported in terms of lateral angle, rather than azimuth. Apart from DWRC, the general trend of the results suggests that source separation performs better in situations where the target is frontal to the listener. This situation was, in fact, one of the classical assumptions made to evaluate source separation methods~\cite{ManWeiEll2010}. By reporting results as in Fig.~\ref{fig:angle_datasets}, we overcome this assumption. The proposed ERIC-MESSL performs better than the others for almost every position of the target source. For the few positions where it is not the best, either the proposed IC-MESSL or ER-MESSL has higher SDRs. In DWRC, the loudspeaker positioned at $27^\circ$ stood next to a chest of drawers, that produces scattering. This conflicts with the overall assumption of having reflections with a dominant specular component. Therefore, the localization of the first reflection, for modeling the comb filter, is affected by estimation errors. Similar to $0^\circ$ in DWRC and $27^\circ$ in Studio1, for $-37^\circ$ in BBC UL, strong lateral reflections arrive before those from the direct sound direction, making the IC dominate the comb filtering effect~\cite{LokPatTerSilSav2011}. Similar results can be observed in Fig.~\ref{fig:angle_datasets_PESQ}, where the PESQ results are reported as a function of the target source location. It is evident how the proposed ERIC-MESSL, which combines the two proposed models, outperforms, in general the baseline MESSL~\cite{ManWeiEll2010}. Furthermore, these PESQ results also show what was already observed in Fig.~\ref{fig:angle_datasets} for the SDRs (and discussed above), ERIC-MESSL mainly suffers when early reflections are not completely specular. 

The majority of the setups that we tested, had a certain configuration that produced, as the first reflection, the one corresponding to the floor (i.e. having same azimuth as the direct sound). Nevertheless, in BBC UL, DWRC, and Studio1, there are cases where the first arriving reflection has a different direction of arrival (DOA) than the direct sound (i.e. coming from a lateral wall). The proposed model does not make any assumption regarding the direction of the reflections, however, the condition that better matches the idea behind it (i.e. a strong comb filter effect) is given by the case of direct sound and early reflection coming from the same direction. 
To better show the strength of the proposed models, in Table~\ref{tab:SDR_1_interferer_same_angle}, we show the results of the experiments by considering only those situations where direct sound and first reflection have the same DOA. These results show that our methods outperform MESSL with a much wider difference than the overall results in Table~\ref{tab:SDR_1_interferer}, and ERIC-MESSL is the best. 

\begin{figure}[!t]
\centering
\includegraphics[trim={40pt 37pt 65pt 0pt}, clip, width=\columnwidth]{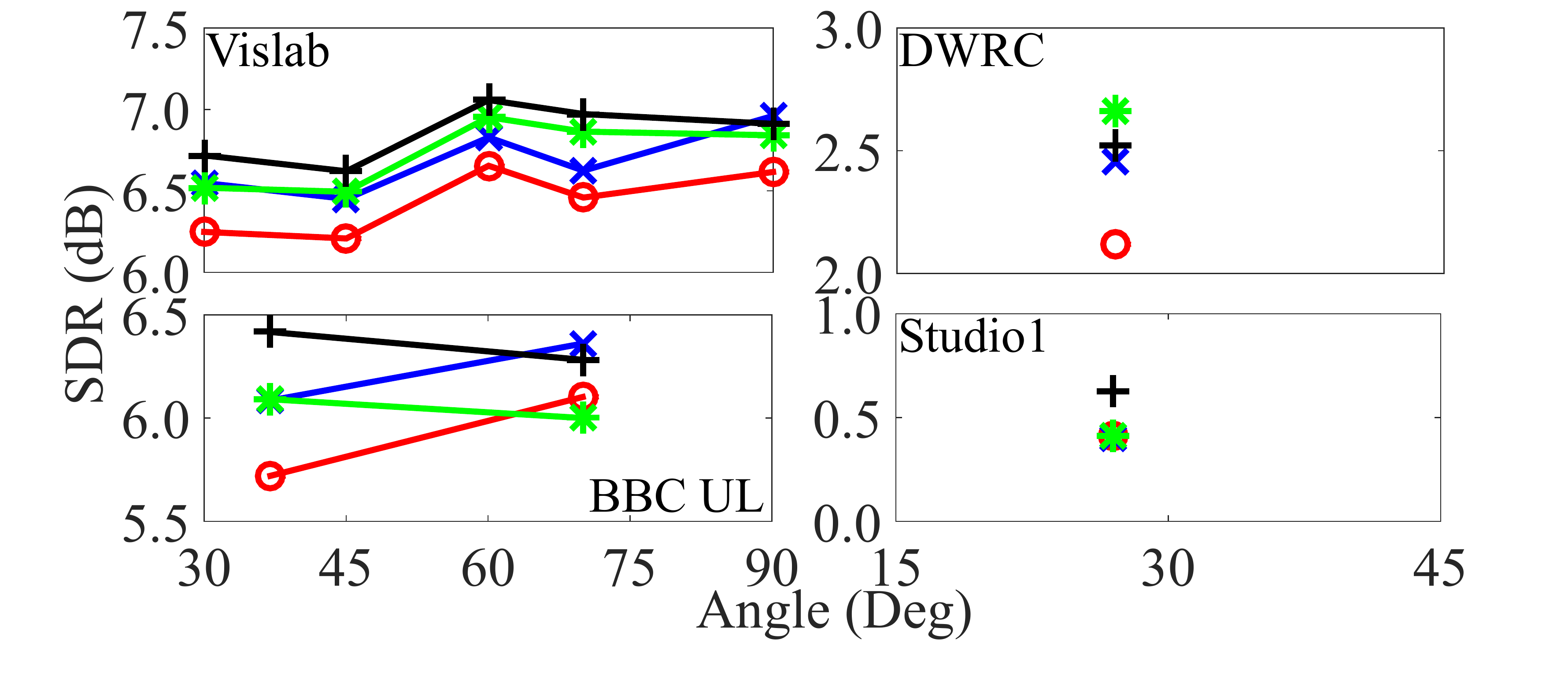}
\caption{SDRs for different interferer positions, fixing target at $0^\circ$. The black vertical crossed lines refer to ERIC-MESSL, the red circled lines to MESSL \cite{ManWeiEll2010}, the green starred lines to ER-MESSL, and the blue crossed lines to IC-MESSL.}
\label{fig:zero_degrees}
\end{figure}

To analyze the effect of separation angle, the source separation performance was  calculated with the frontal loudspeaker ($0^\circ$ azimuth) as the target source, and varying the interferer. The results are reported in Fig. \ref{fig:zero_degrees}, as is typical in the literature for source separation \cite{ManWeiEll2010,SawAraMak2011,AliJacLiuWan2014}. This kind of visualization allows a better understanding of the source separation performance by varying TISA. By observing the results of Vislab and BBC UL (datasets having loudspeaker positions around the listener), the proposed ERIC-MESSL consistently provides the highest performance. However, for the extreme cases of TISA (i.e. 90$^\circ$ in Vislab and 70$^\circ$ in BBC UL), the proposed IC-MESSL performs better. This behavior is best seen in the proposed ER-MESSL results. As for ERIC-MESSL, ER-MESSL is better than IC-MESSL for almost every TISA, apart from the extreme cases (i.e. 90$^\circ$ in Vislab and 70$^\circ$ in BBC UL). Therefore, we can conclude that the comb filter is, on average, more effective than the IC, apart from large TISAs.
For both DWRC and Studio1, all the methods show degradation at low TISA. This is a common source separation problem~\cite{ManWeiEll2010}. Studio1 is also confirmed to be problematic, with SDR lower than 1\,dB, for every method. 

Regarding the overall computational complexity, the average run time, for a code run in MATLAB R2014b on Intel(R) Core(TM)i7-2600 CPU @ 3.40GHz, 16GB RAM PC is 55\,s for ERIC-MESSL and 8\,s for MESSL~\cite{ManWeiEll2010}. The parameters are searched within a 7-D space in ERIC-MESSL, making it less efficient than MESSL, where the space was one dimensional.

\begin{table}[!t]
\caption[]{Evaluation results for the deep learning based methods over Vislab, in terms of SDR, PESQ and ESTOI.}
\label{tab:deep-learning}
\centering
\begin{tabular}{|c|c|c|c|c|c|}
\cline{2-4}
\multicolumn{1}{c|}{}& \textbf{SDR} & \textbf{PESQ} & \textbf{ESTOI} \\
\hline
\textbf{Direct sound information} & 8.33 & 2.51 & 0.70 \\
\hline
\textbf{Direct sound and early reflection info} & $\bm{8.80}$ & $\bm{2.59}$ & $\bm{0.73}$ \\
\hline
\end{tabular}
\end{table}

\textbf{Early Reflections and Deep Learning.}
We now evaluate a DNN-based method that is representative of state-of-the-art approaches in speech separation. We modified this reference method to test the key point behind our main work: that the inclusion of early reflection information into source separation methods improves the performance. This test is intended to examine the potential for exploiting this information using a DNN approach, and give a preliminary validation. Further experiments are needed to explore the best way to incorporate early reflection information within DNN architectures for source separation, beyond the present preliminary integration.

The selected pipeline is based on the classic multilayer perceptron (MLP) architecture, as presented in~\cite{LiuXuJacWanCol2018}. A similar architecture can be also found in~\cite{WanWan2018}. In our implementation, the MLP has two hidden layers, containing 1024 leaky rectified linear units (ReLU) each. We employed batch normalisation (BN) layers~\cite{IofSze2015} to accelerate convergence, and Adam optimizer~\cite{KinBa2015} with He initialization~\cite{HeZhaRenSun2015}. The binary crossentropy was used as loss function. The mini-batch size was set to 1000.

Recordings from two male speakers and two female speakers in the TIMIT dataset~\cite{GarLamFisFisPalDah1993} were used for our experiment. For each of these speakers, ten sentences were randomly selected. The binaural mixtures were generated by convolving the randomly chosen utterances with BRIRs recorded in Vislab. The BRIRs used were the ones recorded for the angles at 0$^\circ$, $\pm30^\circ$ and $\pm60^\circ$. To create the mixtures, each of the 4 speakers was combined to the other 3. For each of these 12 combinations, we associated the 10 sentences. In terms of the product rule for counting, this makes a total of 1200 utterance combinations. Regarding the BRIRs, each of the 5 DOAs was combined to the other 4, making a total of 20 combinations. Convolving utterances with BRIRs, we obtain 24000 mixtures: 19200 were randomly selected for training, the rest for testing.

These 24000 samples comprising the dataset represent all combinations of the BRIR directions convolved with the individual utterances. A distinct set of direction-utterance samples was used for testing and training, although all directions and some utterances did overlap (but not any specific combination). The performance of the methods tested here would likely decrease when generalizing to new unseen utterances and BRIRs, which is however beyond the scope of the present tests. In fact, as mentioned above, this DNN experiment is to demonstrate that, by adding information about early reflections, supervised deep learning based source separation method can also be improved, over the case where only the direct sound is considered, as we observed in the main novelty of this article, i.e. the GMM based unsupervised method.  

The training was performed by providing the features related to the IPD as input to the network, and matching with the ORACLE masks in output. In both models, the IPD features were calculated through the approach in Sections~\ref{secchap:theoretical_definitions} and \ref{secchap:Interaural_Cues_based_Model}. To evaluate the improvement given by the early reflection information, we have trained one model that considers only the direct sound information~\cite{LiuXuJacWanCol2018}, and a novel one which we propose to also incorporate the early reflections. The ORACLE masks in output to the training stage were generated from Equation~(\ref{eq:near_optimal_mask}), by considering $E^{\mathrm{tar}}_l(n,\omega)$ and $E^{\mathrm{int}}_l(n,\omega)$ related to the direct sound for the model used as in~\cite{LiuXuJacWanCol2018}, and direct sound plus early reflections for our model. This was done by segmenting the related BRIRs through a Hamming window (5\,ms, and 30\,ms, respectively).

During the test, the masks predicted by the networks are used to separate the sounds, by employing Equation~(\ref{eq:separated_source}). Results are reported in Table~\ref{tab:deep-learning}. There, it is shown how the model containing information about the early reflections offers better performance with respect to the pipeline which considers only direct sound, for every metric (i.e. SDR, PESQ and ESTOI). This has demonstrated the key idea of the manuscript: early reflections carry important information that is helpful for improving the performance of speech separation models, including both unsupervised (e.g. MESSL) and supervised techniques (e.g. DNNs).
However, it is important to stress that MESSL~\cite{ManWeiEll2010} and the methods proposed in Section~\ref{secchap:source_sep_model} are unsupervised techniques, hence do not need any labeling. Therefore, it is inappropriate to directly compare the results in Table~\ref{tab:deep-learning} with those in Tables~\ref{tab:SDR_1_interferer} and \ref{tab:ESTOI_1_interferer}. 

\section{Conclusion}
\label{secchap:conclusion6}
Two room properties (i.e. early reflections and late reverberation) have been modeled for source separation. Depending on whether they are modeled individually or together, three novel source separation methods have been proposed: ER-MESSL, that models the comb filter effect; IC-MESSL, that models the IC; ERIC-MESSL, that combines the two models together.   

Experiments were performed by recording four reverberant environments, and comparing the source separation performance of the proposed methods with MESSL's~\cite{ManWeiEll2010}. In general, the proposed ERIC-MESSL outperforms all the other methods. With respect to MESSL, the improvement given by ERIC-MESSL, averaged over the four tested datasets, is about 10\,\% for SDR and 2\,\% for PESQ. It was also shown, by running t-tests, that the ERIC-MESSL results are statistically different from MESSL's. Moreover, this experimental analysis revealed that low DRRs and narrow AVG-TISAs led to a degradation of the results. In addition, results were also observed by varying both the target source and interferer positions. Also in this case, it was consistently observed that ERIC-MESSL is, in general, the better model.
We conclude that modeling together the comb filter effect and IC is helpful for improving the performance of classical source separation methods. Furthermore, we have also reported an experiment undertaken by including early reflection information into a DNN based state-of-the-art source separation method. Results showed a great improvement, thus confirming the importance of incorporating the early reflection information into both unsupervised and supervised source separation methods.

Future work may be conducted on extending the methods to multichannel arrays of microphones. Furthermore, a combination of audio-visual sensing may be explored, to tackle problematic scenarios where the interferer has a higher level than the target. The proposed models could also be applied to other popular approaches, such as NMF. 

\setcounter{secnumdepth}{0} 
\section{Acknowledgments}
This work was supported by the EPSRC Programme Grant S3A: Future Spatial Audio for an Immersive Listener Experience at Home (EP/L000539/1) and BBC as part of the BBC Audio Research Partnership. The authors would like to thank the reviewers and the associate editor for their helpful comments to improve the article.

\ifCLASSOPTIONcaptionsoff
  \newpage
\fi

\bibliographystyle{IEEEbib}

\begin{thebibliography}{10}

\bibitem{SutBunSedSedFilTsiBru2010}
A.~Sutin, B.~Bunin, N.~Sedunov, L.~Fillinger, M.~Tsionskiv, and M.~Bruno,
\newblock ``Stevens passive acoustic system for underwater surveillance,''
\newblock in {\em Proc. of the International WaterSide Security Conference},
  Carrara, Italy, 2010.

\bibitem{UngBigStrLaz2004}
M.~Ungureanu, C.~Bigan, R.~Strungaru, and V.~Lazarescu,
\newblock ``Independent component analysis applied in biomedical signal
  processing,''
\newblock {\em Measurement Science Review}, vol. 4, no. 2, pp. 1--8, 2004.

\bibitem{TonSalBed2007}
A.~Tonazzini, E.~Salerno, and L.~Bedini,
\newblock ``Fast correction of bleed-through distortion in grayscale documents
  by a blind source separation technique,''
\newblock {\em International Journal of Document Analysis}, vol. 10, no. 1, pp.
  17--25, 2007.

\bibitem{MohSmaLei2013}
N.~Mohammadiha, P.~Smaragdis, and A.~Leijon,
\newblock ``Supervised and unsupervised speech enhancement using nonnegative
  matrix factorization,''
\newblock {\em IEEE Transactions on Audio, Speech, and Language Processing},
  vol. 21, no. 10, pp. 2140--2151, 2013.

\bibitem{AkeChaBulPalSumNelGat2007}
M.~A. Akeroyd, J.~Chambers, D.~Bullock, Palmer~A. R., and A.~Q. Summerfield,
\newblock ``The binaural performance of a cross-talk cancellation system with
  matched or mismatched setup and playback acoustics,''
\newblock {\em J. Acoustical Society of America}, vol. 121, no. 2, pp.
  1056--1069, 2007.

\bibitem{LiDenGonHae2014}
J.~Li, L.~Deng, Y.~Gong, and R.~Haeb-Umbach,
\newblock ``An overview of noise-robust automatic speech recognition,''
\newblock {\em IEEE/ACM Transactions on Audio, Speech, and Language
  Processing}, vol. 22, no. 4, pp. 745--777, 2014.

\bibitem{HeaYoHWanWan2013}
E.~W. Healy, S.~E. Yoho, Y.~Wang, and Wang D.,
\newblock ``An algorithm to improve speech recognition in noise for
  hearing-impaired listeners,''
\newblock {\em J. Acoustical Society of America}, vol. 134, no. 4, pp.
  3029--3038, 2013.

\bibitem{CroCriTruMur2016}
C.~Crocco, M.~Cristiani, A.~Trucco, and V.~Murino,
\newblock ``Audio surveillance: a systematic review,''
\newblock {\em ACM Computing Surveys}, vol. 48, no. 4, pp. 52:1--52:46, 2016.

\bibitem{LiuWanJacCox2015}
Q.~Liu, W.~Wang, P.~J.~B. Jackson, and T.~J. Cox,
\newblock ``A source separation evaluation method in object-based spatial
  audio,''
\newblock in {\em Proc. of the 23rd European Signal Processing Conference
  (EUSIPCO)}, Nice, France, 2015.

\bibitem{KinDelGanHabHaeKelLeuMaaNakRajSehYos2016}
K.~Kinoshita, M.~Delcroix, S.~Gannot, E.~A.~P. Habets, R.~Haeb-Umbach,
  W.~Kellermann, V.~Leutnant, R.~Maas, T.~Nakatani, B.~Raj, A.~Sehr, and
  T.~Yoshioka,
\newblock ``A summary of the {REVERB} challenge: state-of-the-art and remaining
  challenges in reverberant speech processing research,''
\newblock {\em EURASIP J. on Advances in Signal Processing}, vol. 2016, no. 1,
  pp. 7:1--7:19, 2016.

\bibitem{Kuttruff4}
H.~Kuttruff,
\newblock {\em Room Acoustics - Fifth edition},
\newblock Spon press, 2009.

\bibitem{Blesser2001}
B.~Blesser,
\newblock ``An interdisciplinary synthesis of reverberation viewpoints,''
\newblock {\em J. Audio Engineering Society}, vol. 49, no. 10, pp. 867--903,
  2001.

\bibitem{ValParSavSmiAbe2012}
V.~V\"{a}lim\"{a}ki, J.~A. Parker, L.~Savioja, J.~O. Smith, and J.~S. Abel,
\newblock ``Fifty years of artificial reverberation,''
\newblock {\em IEEE Transactions on Audio, Speech and Language Processing},
  vol. 20, no. 5, pp. 1421--1448, 2012.

\bibitem{Barron1971}
M.~Barron,
\newblock ``The subjective effects of first reflections in concert halls - the
  need for lateral reflections,''
\newblock {\em J. of Sound and Vibration}, vol. 15, no. 4, pp. 475--494, 1971.

\bibitem{LokPatTerSilSav2011}
T.~Lokki, J.~P\"{a}tynen, T.~Sakar, S.~Siltanen, and L.~Savioja,
\newblock ``Engaging concert hall acoustics is made up of temporal envelope
  preserving reflections,''
\newblock {\em J. Acoustical Society of America Express Letters}, vol. 129, no.
  6, pp. EL223--EL228, 2011.

\bibitem{VinBerGriBim2014}
E.~Vincent, N.~Bertin, R.~Gribonval, and F.~Bimbot,
\newblock ``From blind to guided audio source separation,''
\newblock {\em IEEE Signal Processing Magazine}, vol. 31, no. 3, pp. 107--115,
  2014.

\bibitem{ManWeiEll2010}
M.~I. Mandel, R.~J. Weiss, and D.~P.~W. Ellis,
\newblock ``Model-based expectation maximization source separation and
  localization,''
\newblock {\em IEEE Transactions on Audio, Speech, and Language Processing},
  vol. 18, no. 2, pp. 382--394, 2010.

\bibitem{Bech1998}
S.~Bech,
\newblock ``Spatial aspects of reproduced sound in small rooms,''
\newblock {\em J. Acoustical Society of America}, vol. 103, no. 1, pp.
  434--445, 1998.

\bibitem{AliWanJac2013}
A.~Alinaghi, W.~Wang, and P.~J.~B. Jackson,
\newblock ``Spatial and coherence cues based time-frequency masking for
  binaural reverberant speech separation,''
\newblock in {\em Proc. of the IEEE International Conference on Acoustics,
  Speech and Signal Processing (ICASSP)}, Vancouver, Canada, 2013.

\bibitem{RemJacColWan2017}
L.~Remaggi, P.~J.~B. Jackson, P.~Coleman, and W.~Wang,
\newblock ``Acoustic reflector localization: novel image source reversion and
  direct localization methods,''
\newblock {\em IEEE/ACM Transactions on Audio, Speech and Language Processing},
  vol. 25, no. 2, pp. 296--309, 2017.

\bibitem{AllenBerkley79}
J.~B. Allen and D.~A. Berkley,
\newblock ``Image method for efficiently simulating small-room acoustics,''
\newblock {\em J. Acoustical Society of America}, vol. 4, no. 65, pp. 943--950,
  1979.

\bibitem{JanLee2003}
G-J. Jang and T-W. Lee,
\newblock ``A maximum likelihood approach to single-channel source
  separation,''
\newblock {\em J. of Machine Learning Research}, vol. 23, pp. 1365--1392, 2003.

\bibitem{SchOls2006}
M.~N. Schmidt and R.~K. Olsson,
\newblock ``Single-channel speech separation using sparse non-negative matrix
  factorization,''
\newblock in {\em Proc. of Interspeech}, Pittsburgh, USA, 2006.

\bibitem{ArbOzeDuoVinGriBimVan2010}
S.~Arberet, A.~Ozerov, N.~Q.~K. Duong, E.~Vincent, R.~Gribonval, F.~Bimbot, and
  P.~Vandergheynst,
\newblock ``Nonnegative matrix factorization and spatial covariance model for
  under-determined reverberant audio source separation,''
\newblock in {\em Proc. of the 10th International Conference on Information
  Science, Signal Processing and their Applications (ISSPA)}, Kuala Lumpur,
  Malaysia, 2010.

\bibitem{JodWenEybVirSch2012}
C.~Joder, F.~Weninger, F.~Eyben, D.~Virette, and B.~Schuller,
\newblock ``Real-time speech separation by semi-supervised nonnegative matrix
  factorization,''
\newblock in {\em Latent Variable Analysis and Signal Separation: 10th
  International Conference (LVA/ICA)}. Tel Aviv, Israel, 2012, pp. 322--329,
  Springer Berlin Heidelberg.

\bibitem{SmaFevMysMohHof2014}
P.~Smaragdis, C.~F\'{e}votte, G.~J. Mysore, N.~Mohammadiha, and M.~Hoffman,
\newblock ``Static and dynamic source separation using nonnegative
  factorizations: A unified view,''
\newblock {\em IEEE Signal Processing Magazine}, vol. 31, no. 3, pp. 66--75,
  2014.

\bibitem{SawAraMukMak2006}
H.~Sawada, S.~Araki, R.~Mukai, and S.~Makino,
\newblock ``Blind extraction of dominant target sources using {ICA} and
  time-frequency masking,''
\newblock {\em IEEE Transactions on Audio, Speech and Language Processing},
  vol. 14, no. 6, pp. 2165--2173, 2006.

\bibitem{OzeFev2010}
A.~Ozerov and C.~F\'{e}votte,
\newblock ``Multichannel nonegative matrix factorization in convolutive
  mixtures for audio source separation,''
\newblock {\em IEEE Transactions on Audio, Speech and Language Processing},
  vol. 18, no. 3, pp. 550--563, 2010.

\bibitem{SouAraKinNakSaw2013}
M.~Souden, S.~Araki, K.~Kinoshita, T.~Nakatani, and H.~Sawada,
\newblock ``A multichannel {MMSE}-based framework for speech source separation
  and noise reduction,''
\newblock {\em IEEE Transactions on Audio, Speech, and Language Processing},
  vol. 21, no. 9, pp. 1913--1928, 2013.

\bibitem{WanReiCav2016}
L.~Wang, J.~D. Reiss, and A.~Cavallaro,
\newblock ``Over-determined source separation and localization using
  distributed microphones,''
\newblock {\em IEEE/ACM Transactions on Audio, Speech and Language Processing},
  vol. 24, no. 9, pp. 1573--1588, 2016.

\bibitem{GanVinMarGolOze2017}
S.~Gannot, E.~Vincent, S.~Markovich-Golan, and A.~Ozerov,
\newblock ``A consolidated perspective on multimicrophone speech enhancement
  and source separation,''
\newblock {\em IEEE/ACM Transactions on Audio, Speech and Language Processing},
  vol. 25, no. 4, pp. 692--730, 2017.

\bibitem{SarKawNisLeeShi2006}
H.~Saruwatari, T.~Kawamura, T.~Nishikawa, A.~Lee, and K.~Shikano,
\newblock ``Blind source separation based on a fast-convergence algorithm
  combining {ICA} and beamforming,''
\newblock {\em IEEE Transactions on Audio, Speech and Language Processing},
  vol. 14, no. 6, pp. 2165--2173, 2006.

\bibitem{DokSchVet2015}
I.~Dokmani\'{c}, R.~Scheibler, and M.~Vetterli,
\newblock ``Raking the cocktail party,''
\newblock {\em IEEE J. of Selected Topics in Signal Processing}, vol. 9, no. 5,
  pp. 825--836, 2015.

\bibitem{HuaKimHasJohSma2015}
P.-S. Huang, M.~Kim, M.~Hasegawa-Johnson, and P.~Smaragdis,
\newblock ``Joint optimization of masks and deep recurrent neural networks for
  monaural source separation,''
\newblock {\em IEEE/ACM Transactions on Audio, Speech and Language Processing},
  vol. 23, no. 12, pp. 2136--2147, 2015.

\bibitem{NugLiuVin2016}
A.~A. Nugraha, A.~Liutkus, and E.~Vincent,
\newblock ``Multichannel audio source separation with deep neural networks,''
\newblock {\em IEEE/ACM Transactions on Audio, Speech and Language Processing},
  vol. 24, no. 9, pp. 1652--1664, 2016.

\bibitem{ZhaWan2016}
X.-L. Zhang and D.~L. Wang,
\newblock ``A deep ensemble learning method for monaural speech separation,''
\newblock {\em IEEE/ACM Transactions on Audio, Speech and Language Processing},
  vol. 24, no. 5, pp. 967--977, 2016.

\bibitem{DuTuDa2016}
J.~Du, Y.~Tu, L-R. Dai, and C.-H. Lee,
\newblock ``A regression approach to single-channel speech separation via
  high-resolution deep neural networks,''
\newblock {\em IEEE/ACM Transactions on Audio, Speech and Language Processing},
  vol. 24, no. 8, pp. 1424--1437, 2016.

\bibitem{WanDuDai2017}
Y.~Wang, J.~Du, L.-R. Dai, and C.-H. Lee,
\newblock ``A gender mixture detection approach to unsupervised single-channel
  speech separation based on deep neural networks,''
\newblock {\em IEEE/ACM Transactions on Audio, Speech and Language Processing},
  vol. 25, no. 7, pp. 1535--1546, 2017.

\bibitem{Wang2008}
D.~Wang,
\newblock ``Time-frequency masking for speech separation and its potential for
  hearing aid design,''
\newblock {\em Trends in Amplification}, vol. 12, no. 4, pp. 332--353, 2008.

\bibitem{HofVan1998}
P.~M. Hofman and J.~Van~Opstal,
\newblock ``Spectro-temporal factors in two-dimensional human sound
  localization,''
\newblock {\em J. Acoustical Society of America}, vol. 103, no. 5, pp.
  2634--2648, 1998.

\bibitem{SawAraMak2011}
H.~Sawada, S.~Araki, and S.~Makino,
\newblock ``Underdetermined convolutive blind source separation via frequency
  bin-wise clustering and permutation alignment,''
\newblock {\em IEEE Transactions on Audio, Speech, and Language Processing},
  vol. 19, no. 3, pp. 516--527, 2011.

\bibitem{AliJacLiuWan2014}
A.~Alinaghi, P.~J.~B. Jackson, Q.~Liu, and W.~Wang,
\newblock ``Joint mixing vector and binaural model based stereo source
  separation,''
\newblock {\em IEEE/ACM Transactions on Audio, Speech and Language Processing},
  vol. 22, no. 9, pp. 1434--1448, 2014.

\bibitem{DelForHor2015}
A.~Deleforge, F.~Forbes, and R.~Horaud,
\newblock ``Acoustic space learning for sound-source separation and
  localization on binaural manifolds,''
\newblock {\em International Journal of Neural Systems}, vol. 25, no. 1, 2015.

\bibitem{HumMasBro2010}
C.~Hummersone, R.~Mason, and T.~Brookes,
\newblock ``Dynamic precedence effect modeling for source separation in
  reverberant environments,''
\newblock {\em IEEE Transactions on Audio, Speech, and Language Processing},
  vol. 18, no. 7, pp. 1867--1871, 2010.

\bibitem{HuaBenChe2005}
Y.~Huang, J.~Benesty, and J.~Chen,
\newblock ``A blind channel identification-based two-stage approach to
  separation and dereverberation of speech signals in a reverberant
  environment,''
\newblock {\em IEEE Transactions on Audio, Speech and Language Processing},
  vol. 13, no. 5, pp. 882--895, 2005.

\bibitem{NesOmo2012}
F.~Nesta and M.~Omologo,
\newblock ``Convolutive underdetermined source separation through weighted
  interleaved {ICA} and spatio-temporal source correlation,''
\newblock in {\em Latent Variable Analysis and Signal Separation: 10th
  International Conference (LVA/ICA)}. Tel Aviv, Israel, 2012, pp. 222--230,
  Springer Berlin Heidelberg.

\bibitem{MakSawLee2007}
S.~Makino, H.~Sawada, and T.~W. Lee,
\newblock {\em Blind Speech Separation},
\newblock Springer, 2007.

\bibitem{AsaGolBouCev2014}
A.~Asaei, M.~Golbabaee, H.~Bourlard, and V.~Cevher,
\newblock ``Structured sparsity models for reverberant speech separation,''
\newblock {\em IEEE/ACM Transactions on Audio, Speech and Language Processing},
  vol. 22, no. 30, pp. 620--633, 2014.

\bibitem{SchDiCDelDok2017}
R.~Scheibler, D.~Di Carlo, A.~Deleforge, and I.~Dokmanic,
\newblock ``Separake: Source separation with a little help from echoes,''
\newblock {\em arXiv: CoRR}, vol. abs/1711.06805, 2017.

\bibitem{VinVirGan2018}
E.~Vincent, T.~Virtanen, and S.~Gannot,
\newblock {\em Audio source separation and speech enhancement},
\newblock John Wiley \& Sons, Ltd, 2018.

\bibitem{BroCoo1994}
G.~J. Brown and M.~Cook,
\newblock ``Computational auditory scene analysis,''
\newblock {\em Computer Speech and Language}, vol. 8, pp. 297--336, 1994.

\bibitem{ValMicRou2007}
J.-M. Valin, F.~Michaud, and J.~Rouat,
\newblock ``Robust localization and tracking of simultaneous moving sound
  sources using beamforming and particle filtering,''
\newblock {\em Robotics and Autonomous Systems}, vol. 55, no. 1, pp. 216--228,
  2007.

\bibitem{NaqYuCha2010}
S.~M. Naqvi, M.~Yu, and J.~A. Chambers,
\newblock ``A multimodal approach to blind source separation of moving
  sources,''
\newblock {\em IEEE Journal of Selected Topics in Signal Processing}, vol. 4,
  no. 5, pp. 895--910, 2010.

\bibitem{FalMer2004}
C.~Faller and J.~Merimaa,
\newblock ``Source localization in complex listening situations: Selection of
  binaural cues based on interaural coherence,''
\newblock {\em The Journal of the Acoustical Society of America}, vol. 116, no.
  5, pp. 3075--3089, 2004.

\bibitem{JeuSchEscVar2010}
M.~Jeub, M.~Sch\"{a}fer, T.~Esch, and P.~Vary,
\newblock ``Model-based dereverberation preserving binaural cues,''
\newblock {\em IEEE Transactions on Audio, Speech, and Language Processing},
  vol. 18, no. 7, pp. 1732--1745, 2010.

\bibitem{Aarabi2002}
P.~Aarabi,
\newblock ``Self-localizing dynamic microphone arrays,''
\newblock {\em IEEE Transactions on Systems, Man, and Cybernetics, Part C
  (Applications and Reviews)}, vol. 32, no. 4, pp. 474--484, 2002.

\bibitem{KimRemJacFazHil2017}
H.~Kim, L.~Remaggi, P.~J.~B. Jackson, F.~M. Fazi, and A.~Hilton,
\newblock ``{3D} room geometry reconstruction using audio-visual sensors,''
\newblock in {\em Proc. of the Conference on 3D Vision (3DV)}, Qingdao, China,
  2017.

\bibitem{VanVeenBuck1988}
B.~D. VanVeen and K.~M. Buckley,
\newblock ``Beamforming: a versatile approach to spatial filtering,''
\newblock {\em IEEE Acoustic, Speech and Signal Processing Magazine}, vol. 5,
  no. 2, pp. 4--24, 1988.

\bibitem{Farina2000}
A.~Farina,
\newblock ``Simultaneous measurement of impulse response and distortion with a
  swept-sine technique,''
\newblock in {\em Proc. of the 108th Audio Engineering Society Convention
  (AES)}, Paris, France, 2000.

\bibitem{Zahorik2002}
P.~Zahorik,
\newblock ``Direct-to-reverberant energy ratio sensitivity,''
\newblock {\em J. Acoustical Society of America}, vol. 112, no. 5, Pt. 1, pp.
  2110--2117, 2002.

\bibitem{GarLamFisFisPalDah1993}
J.~S. Garofolo, L.~F. Lamel, W.~M. Fisher, J.~G. Fiscus, D.~S. Pallet, and
  N.~L. Dahlgren,
\newblock ``{DARPA} {TIMIT} acoustic phonetic continuous speech corpus
  {CDROM},''
\newblock Tech. {R}ep., NIST Interagency, 1993.

\bibitem{VinGriFev2006}
E.~Vincent, R.~Gribonval, and C.~F\'{e}votte,
\newblock ``Performance measurement in blind audio source separation,''
\newblock {\em IEEE Transactions on Audio, Speech and Language Processing},
  vol. 14, no. 4, pp. 1462--1469, 2006.

\bibitem{Loizou2013}
P.~C. Loizou,
\newblock {\em Speech Enhancement: Theory and Practice - Second Edition},
\newblock CRC Press, 2013.

\bibitem{JenTaa2016}
J.~Jensen and C.~H. Taal,
\newblock ``An algorithm for predicting the intelligibility of speech masked by
  modulated noise maskers,''
\newblock {\em IEEE/ACM Transactions on Audio, Speech and Language Processing},
  vol. 24, no. 11, pp. 2009--2022, 2016.

\bibitem{VinAraTheNolBofSawOzeGowLutDuo2012}
E.~Vincent, S.~Araki, F.~Theis, G.~Nolte, P.~Bofill, H.~Sawada, A.~Ozerov,
  V.~Gowreesunker, D.~Lutter, and N.~Q.~K. Duong,
\newblock ``The signal separation evaluation campaign (2007-2010): achievements
  and remaining challenges,''
\newblock {\em Signal Processing}, vol. 92, no. 8, pp. 1928--1936, 2012.

\bibitem{Wang2005}
D.~Wang,
\newblock ``On ideal binary mask as the computational goal of auditory scene
  analysis,''
\newblock in {\em Speech Separation by Humans and Machines}, P.~Divenyi, Ed.,
  chapter~12, pp. 181--197. Kluwer Academic, 2005.

\bibitem{WenArrKisWig1993}
E.~M. Wenzel, M.~Arruda, D.~J. Kistler, and F.~L. Wightman,
\newblock ``Localization using nonindividualized head-related transfer
  functions,''
\newblock {\em J. Acoustical Society of America}, vol. 94, no. 1, pp. 111--123,
  1993.

\bibitem{LiuXuJacWanCol2018}
Q.~Liu, Y.~Xu, P.~J.~B. Jackson, W.~Wang, and P.~Coleman,
\newblock ``Iterative deep neural networks for speaker-independent binaural
  blind speech separation,''
\newblock in {\em Proc. of the IEEE International Conference on Acoustics,
  Speech and Signal Processing (ICASSP)}, Brisbane, Canada, 2018.

\bibitem{WanWan2018}
Z.-Q. Wang and D.~Wang,
\newblock ``On spatial features for supervised speech separation and its
  application to beamforming and robust {ASR},''
\newblock in {\em Proc. of the IEEE International Conference on Acoustics,
  Speech and Signal Processing (ICASSP)}, Brisbane, Canada, 2018.

\bibitem{IofSze2015}
S.~Ioffe and C.~Szegedy,
\newblock ``Batch normalization: Accelerating deep network training by reducing
  internal covariate shift,''
\newblock in {\em Proc. of the International Conference on Machine Learning},
  Lille, France, 2015.

\bibitem{KinBa2015}
D.~P. Kingma and J.~L. Ba,
\newblock ``{ADAM}: A method for stochastic optimization,''
\newblock in {\em Proc. of the International Conference on Learning
  Representations (ICLR)}, San Diego, USA, 2015.

\bibitem{HeZhaRenSun2015}
K.~He, X.~Zhang, S.~Ren, and J.~Sun,
\newblock ``Delving deep into rectifiers: surpassing human-level performance on
  imagenet classification,''
\newblock in {\em Proc. of the International Conference on Computer Vision
  (ICCV)}, Santiago, Chile, 2015.

\end{thebibliography}

\vspace*{-1.2cm}
\begin{IEEEbiography}[{\includegraphics[height=1.25in,clip,keepaspectratio]{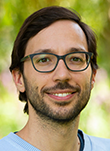}}]{Luca Remaggi} is Audio Research Engineer at Creative Labs, UK, working on cutting edge spatial audio products. Between 2017 and 2019, he was Research Fellow at the Centre for Vision, Speech and Signal Processing, University of Surrey, UK, where he also pursued his PhD, in 2017. His research interest was to investigate the multipath sound propagation combining acoustic and visual data, for applications in spatial audio and source separation. He received the B.Sc. and M.E. degrees in Electronic Engineering from Universit\`{a} Politecnica delle Marche, Italy, in 2009 and 2012, respectively. During his M.E., he has been an intern at the Department of Signal Processing and Acoustics, Aalto University, Finland, where he focused on the sound synthesis of musical instruments. 
\end{IEEEbiography}
\vspace*{-1.2cm}
\begin{IEEEbiography}[{\includegraphics[height=1.25in,clip,keepaspectratio]{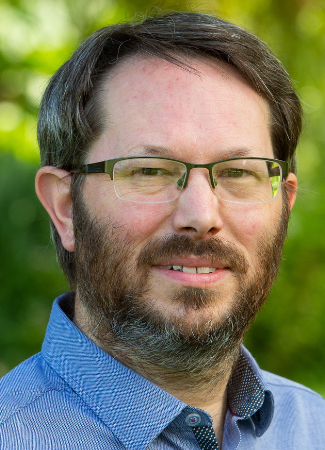}}]{Philip Jackson}
is Reader in Machine Audition at the Centre for Vision, Speech \& Signal Processing (CVSSP, University of Surrey, UK) with MA in Engineering (Cambridge University, UK) and PhD in Electronic Engineering (University of Southampton, UK). His broad interests in acoustical signals have led to research contributions in sound field control, modeling speech articulation, acoustics and recognition, in audio-visual perception, blind source separation, and spatial audio reverberation, capture, reproduction and quality evaluation [h-index 22; Google Scholar: bit.ly/2oTRw1C]. He led one of four research streams on object-based spatial audio in the S3A programme grant funded in the UK by EPSRC, and enjoys listening.
\vspace*{-1.2cm}
\end{IEEEbiography}
\begin{IEEEbiography}[{\includegraphics[height=1.25in,clip,keepaspectratio]{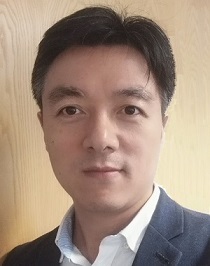}}]{Wenwu Wang}
(M’02–SM’11) was born in Anhui, China. He received the B.Sc. degree in 1997, the M.E. degree in 2000, and the Ph.D. degree in 2002, all from Harbin Engineering University, China. 
He then worked in King’s College London (2002-2003), Cardiff University (2004-2005), Tao Group Ltd. (now Antix Labs Ltd.) (2005-2006), and Creative Labs (2006-2007), before joining University of Surrey, UK, in May 2007, where he is currently a Professor in Signal Processing and Machine Learning, and a Co-Director of the Machine Audition Lab within the Centre for Vision Speech and Signal Processing. He was a Visiting Scholar at Ohio State University, USA, in 2008. He has been a Guest Professor on Machine Perception at Qingdao University of Science and Technology, China, since 2018. 
His current research interests include blind signal processing, sparse signal processing, audio-visual signal processing, machine learning and perception, artificial intelligence, machine audition (listening), and statistical anomaly detection. He has (co)-authored over 250 publications in these areas. 
He and his team have won the Best Paper Award on LVA/ICA 2018, the Best Oral Presentation on FSDM 2016, the Top Paper Award in IEEE ICME 2015, Best Student Paper Award shortlists on IEEE ICASSP 2019 and LVA/ICA 2010. His papers are among the Most Downloaded Papers in IEEE/ACM Transactions on Audio Speech and Language Processing in 2018 and 2019, and Featured Articles in IEEE Transactions on Signal Processing 2013. As a team member, he achieved the 2nd place (among 23 teams) in the DCASE 2019 Challenge “Sound event localization and detection”, the 3rd place (among 558 submitted systems) in the 2018 Kaggle Challenge "Free-sound general purpose audio tagging", the 1st place (among 35 submitted systems) in the 2017 DCASE Challenge on "Large-scale weakly supervised sound event detection for smart cars", the TVB Europe Award for Best Achievement in Sound in 2016 and the finalist for GooglePlay Best VR Experience in 2017, and the Best Solution Award on the Dstl Challenge "Under-sampled signal signal recognition" in 2012. 
He is a Senior Area Editor (2019-) for IEEE Transactions on Signal Processing and an Associate Editor (2019-) for EURASIP Journal on Audio Speech and Music Processing. He was an Associate Editor (2014-2018) for IEEE Transactions on Signal Processing. He was a Publication Co-Chair for ICASSP 2019, Brighton, UK.
\end{IEEEbiography}

\end{document}